\documentclass[english,12pt,sort&compress]{iopart}
\usepackage{setstack}
\usepackage{iopams}
\usepackage{float}
\usepackage{graphicx}
\usepackage{amssymb}
\usepackage{verbatim}
\usepackage{booktabs}
\usepackage{array}
\usepackage[numbers,merge]{natbib}

\usepackage{babel}
\usepackage{subfig}

\newcommand{\fig}{Fig.\ }
\newcommand{\sect}{Sec.\ }
\newcommand{\eqn}{Eq.\ }
\newcommand{\eqns}{Eqs.\ }

\newcommand{\myxrightarrow}[2][{}]{\overset{#2}{\longrightarrow}}
\newcommand{\mytext}[1]{{\mbox{#1}}}
\newcommand{\mysubtext}[1]{{\mathrm{#1}}}

\begin{document}


\title{Motion of condensates in non-Markovian zero-range dynamics}

\author{Ori Hirschberg$^1$, David Mukamel$^1$, Gunter M. Sch\"utz$^2$}

\address{$^1$ Department of Physics of Complex Systems, Weizmann
Institute of Science, Rehovot 76100, Israel}
\address{$^2$ Theoretical Soft Matter and Biophysics, Institute of
Complex Systems, Forschungszentrum J\"ulich, 52425 J\"ulich,
Germany}

\ead{\mailto{ori.hirschberg@weizmann.ac.il},
\mailto{david.mukamel@weizmann.ac.il},
\mailto{g.schuetz@fz-juelich.de}}

\begin{abstract}
Condensation transition in a non-Markovian zero-range process is
studied in one and higher dimensions. In the mean-field
approximation, corresponding to infinite range hopping, the model
exhibits condensation with a stationary condensate, as in the
Markovian case, but with a modified phase diagram. In the case of
nearest-neighbor hopping, the condensate is found to drift by a
``slinky'' motion from one site to the next. The mechanism of the
drift is explored numerically in detail. A modified model with
nearest-neighbor hopping which allows exact calculation of the
steady state is introduced. The steady state of this model is found
to be a product measure, and the condensate is stationary.\\
\end{abstract}

\date{\today}


\section{Introduction}\label{sec:introduction}
In recent years, much progress has been made in the theoretical
understanding of nonequilibrium condensation
\cite{evanszrpreview,MajumdarCondensationLesHouches,Schadschneider2010Book}.
Condensation phenomena of this type are known to occur in a large
variety of systems, where a macroscopic fraction of a conserved
``mass'' is accumulated in a microscopic portion of an extended
system. Some examples of such systems are compartmentalized shaken
granular gases \cite{vanderweele2001}, gelation, i.e., the formation
of a macroscopic hub in complex network
\cite{networksevolutionbook}, condensation of wealth in economics
\cite{BurdaEtal2002WealthZRP}, and the formation of traffic jams on
highways \cite{kaupuzsetal2005zrptraffic}.

An elucidation of the mechanism which lies behind the condensation
transition in such systems was achieved by studying simplified but
prototypical toy models, most notably the zero-range process (ZRP).
In this process, particles hop stochastically between boxes with
hopping rates which depend only on the occupation of the box from
which each particle departs. This models diffusing particles which
``interact'' only with particles at the same location (i.e., the
same box), and hence the name of the process. For any choice of
hopping rates, the steady-state distribution of particles is known
to factorize into single box terms and thus it may be computed
exactly. Using the exact solution, it was shown in
\cite{EvansZRPcondensation} that a condensation transition may occur
when the rate of hopping out of a box decreases with its occupation
(modeling an attractive ``interaction''). In this condensation
transition, the system is homogeneous at small particle densities,
while when the density exceeds a critical value, a single (randomly
chosen) site is occupied by a macroscopic fraction of all particles.

When studying systems which exhibit a condensation transition, such
as those listed above, the ZRP may be used to gain qualitative, and
sometimes also quantitative, insight. This is done by mapping the
dynamics of the system under consideration, usually in an
approximate way, to that of the ZRP, and then utilizing known
results for the ZRP
\cite{kafrietal2002criterion,ProbeParticles2005,ProbeParticlesDynamics2006,ChatterjeeBarma2007,ChatterjeeBarma2008}.
Such a mapping is achieved by disregarding some of the structure of
the original system, so that it can be reduces to the simplified
balls and boxes picture of the ZRP. Underlying this procedure is an
assumption that the condensation transition in the ZRP is universal
in some sense, i.e., that the details which are lost when mapping a
system to the ZRP are irrelevant to condensation.

This universality assumption was recently examined by studying how
disregarding dynamical degrees of freedom may affect the
condensation transition \cite{HirschbergEtal2009}. There, a small
variation of the ZRP, which introduces temporal correlations in its
dynamics, was found to have two main effects on the condensation
transition: (i) a calculation in a mean-field setting showed that
the non-Markovian nature of the dynamics ``renormalizes'' the
parameters and critical exponents of the ZRP, and (ii) a numerical
study of the model on a one-dimensional ring revealed that the
temporally-correlated dynamics induces a ``slinky'' motion of the
condensate throughout the system, whereby the condensate spills over
from one site to the next. Thus, the nature of the condensed phase
is modified in a qualitative manner.

In this paper we present a detailed analysis of the model introduced
in \cite{HirschbergEtal2009}. First, we elaborate on the mean-field
solution which was presented in \cite{HirschbergEtal2009} and
generalize the results to a much broader class of hopping rates. We
then turn to dynamics on a ring with hopping to the nearest-neighbor
site and present a detailed study of the mechanism for the
condensate drift. In studying the behavior of finite rings (of size
$L\leq 2000$ sites), we identify two different modes of condensate
motion: motion through a barrier, in which the condensate is carried
by a single site and spilling to the next site is initiated only
once it overcomes a barrier, and motion with no barrier, in which
the condensate is in a continual motion.

In addition, we consider a somewhat modified non-Markovian dynamics
which allows an exact computation of the steady state on lattices of
any dimension, even in the case of nearest-neighbor hopping. For
this variant of the model we show that the steady-state distribution
is given by a product measure, like in the Markovian ZRP, even
though the currents of particles are temporally correlated. The
exact solution of the model is shown to be qualitatively similar to
the mean-field calculation, and in particular there is no condensate
drift.

The paper is organized as follows. After describing the
non-Markovian ZRP in \sect \ref{sec:description}, we present in
\sect \ref{sec:MeanField} the full mean-field solution of the model
and study how condensation is affected by the non-Markovian
dynamics. The numerical study of the condensation transition on a
lattice with nearest-neighbor hopping is presented in \sect
\ref{sec:Ring}. There we examine a particular choice of rates, which
we term the on-off model, on a symmetric, totally asymmetric and
partially asymmetric dynamics, and we analyze the mechanism for the
motion of the condensate. In \sect \ref{sec:Exact}, we present the
exactly solvable variant of the non-Markovian ZRP, calculate its
steady state distribution and discuss condensation in this model.


\section{Description of the model}\label{sec:description}
We consider a system of $N$ particles hopping between $L$ boxes
(labeled by $i = 1,\ldots,L$) with a mean density $\rho = N/L$. We
are mainly interested in the thermodynamic limit in which $L,N \to
\infty$ with the density $\rho$ kept constant. The state of each box
is given by two variables: the number of particles in the box $n_i$,
and a ``clock'' variable $\tau_i$. Both variable take non-negative
integer values: $n_i,\tau_i = 0,1,2,\ldots$. A configuration of the
system is thus given by the set of pairs $({\mathbf n},
\boldsymbol{\tau}) = \{(n_i,\tau_i)\}_{i=1}^L$.

The dynamics proceeds by particles jumping one at a time between the
boxes, and, in parallel, by advances of the clocks. Conforming with
the ``zero-range'' character of the ZRP, the rate with which a
particle leaves a box is taken to depend only on the state of the
box, i.e., on its occupation and clock state. We denote these
hopping rates by $u(n,\tau)$. The clock dynamics is correlated with
particle jumps: every time a particle jumps into a box, its clock is
reset to zero. Independently, the clocks are advanced with a
constant probability per unit time $c$. The two types of dynamical
moves (a jump of a particle between two boxes $i$ and $j$, and and
an advance of the clock at site $i$) can be written as
\begin{eqnarray}\label{eq:modeldscrp}
(n_i,\tau_i),(n_{j},\tau_{j})&\overset{p_{ij}u(n_i,\tau_i)}{\longrightarrow}&
(n_i\!-\!1,\tau_i),(n_{j}\!+\!1,\tau_j=0)
\nonumber \\
(n_i,\tau_i)&\overset{c}{\longrightarrow}& (n_i,\tau_i\!+\!1).
\end{eqnarray}
Here, $p_{ij}$ is a connectivity matrix which states the probability
that a particle departing from site $i$ will choose site $j$ as its
target. Particular choices of $p_{ij}$ are further discussed below.
The jump rates must satisfy $u(0,\tau)=0$ for all $\tau$, as a
particle cannot jump out of an empty box.

To be precise about the meaning of ``rate'' this dynamics can be
rephrased as follows: each box $i$ of the lattice carries two
alarm-clocks which ring after some random time. All $2L$
alarm-clocks ring independently. Given that the state at box $i$ is
$(n_i,\tau_i)$, alarm-clock number 1 of this site rings after an
exponentially distributed random time with parameter $u(n,\tau)$,
while alarm-clock number 2 rings after an exponentially distributed
random time with parameter $c$. If clock 1 rings first, a particle
selects a target box $j$ with probability $p_{ij}$ and jumps to it.
If clock 2 rings first, the internal clock at box $i$ is incremented
by one unit. After any change of the state at box $i$, the clocks
ring again after an exponentially distributed random time determined
by the updated state at box $i$.

The dynamical rules (\ref{eq:modeldscrp}) are written for a general
connectivity matrix $p_{ij}$. In this paper we concentrate mainly on
two schemes for the choice of target box: mean-field (MF) dynamics,
and a one-dimensional homogeneous ring geometry with
nearest-neighbor hopping. In mean-field (MF) dynamics, the target
box $j\neq i$ is chosen randomly and uniformly from all boxes, i.e.,
\begin{equation}
p_{ij} = \frac{1}{L-1} \quad \mytext{for all $j \neq i$}.
\end{equation}
In the case of ring dynamics, on the other hand, particles hop only
between nearest-neighbor sites, possibly in an asymmetric fashion.
Accordingly, the target box is chosen to be $i+1$ with probability
$1-p$ and $i-1$ with probability $p$ (where box $L+1$ is identified
with box 1), i.e.,
\begin{equation}\label{eq:ZRPringdynamics}
p_{ij} = \left\{\begin{array}{ll} (1-p) & \mytext{(if $j=i+1$)}\\
p  & \mytext{(if $j=i-1$)} \\
0 & \mytext{otherwise} \end{array} \right..
\end{equation}
Here, $0\leq p \leq 1$ is the asymmetry parameter: when it is equal
to 0 the hopping is totally asymmetric and no hopping backwards can
occur, while when $p=1/2$ the dynamics is completely symmetric.
Below, MF dynamics is studied in \sect \ref{sec:MeanField}, while
ring dynamics is studied in Sections \ref{sec:Ring} and
\ref{sec:Exact}. We also briefly examine (in Sections
\ref{sec:RingHigherD} and \ref{sec:ExactSym}) dynamics on
higher-dimensional lattices. The generalization of
(\ref{eq:ZRPringdynamics}) to the higher-dimensional case is rather
straightforward and will be presented below when it is discussed.

The dependence of the jump rates on $\boldsymbol{\tau}$ renders the
particle jump process, when it is taken by itself, non-Markovian, as
the rate of a jump depends on how much time has passed since a
particle hopped into the jump site. When considered in the higher
dimensional space of occupations together with clocks, the full
jump/increment process defined above is Markovian. Nevertheless, we
will refer to this process as the non-Markovian ZRP, to stress the
history dependence of the jump process.

The process may be implemented by a discrete-time Monte-Carlo
version of this dynamics with random sequential update which is
defined as follows: define $p_{\max} = \max_{n,\tau} [u(n,\tau) +  c
]$. For the Monte-Carlo update pick a random box uniformly and
attempt to make one of the following changes: (i) move a particle to
a target box (selected according to the appropriate scheme) with
probability $u(n,\tau)/p_{\max}$, (ii) increment the internal clock
with probability $c/p_{\max}$. A total of $Lp_{\max}$ consecutive
update attempts constitute one Monte-Carlo time unit.

A final remark on the nature of the clock variables. As is clear
from the dynamical rules, the clock variables proceed in an
irregular, stochastic fashion. Therefore, they do not measure the
exact time that has passed since a particle last entered each site.
Choosing clock variables which really measure time, i.e., which are
continuous and proceed regularly, might seem more natural for some
physical applications. Such regular clocks are not considered below,
but we remark that they may be achieved starting from the dynamics
(\ref{eq:modeldscrp}) by taking an appropriate limit. This procedure
is described in \ref{appendix:RegClocks}.

\section{The non-Markovian ZRP with mean-field
dynamics}\label{sec:MeanField}
\subsection{General observations}
In this section we analyze the non-Markovian ZRP with mean-field
dynamics. In particular, we investigate how condensation is affected
by the non-Markovian nature of the jump process.

Since each jump of a particle is to an arbitrarily chosen box, the
MF dynamics does not generate correlations between different boxes
beyond the correlations which arise from conservation of particles.
Therefore, in the thermodynamic limit, the stationary distribution
is expected to factorize into a product of single-box terms
\begin{equation}\label{eq:mfZRPproductmeasure}
\mathcal{P}({\mathbf n},\boldsymbol{\tau}) =
Z^{-1}_{L,N}\prod_{i=1}^L P(n_i,\tau_i)\,\delta\bigg(\!\sum_{i=1}^L
n_i - N\bigg),
\end{equation}
where $P(n_i,\tau_i)$ are the single-box occupation and clock
probabilities, and the $\delta$ function is a consequence of the
conservation of particles. The normalization is given by
\begin{equation}\label{eq:mfZRPpartition}
Z_{L,N} = \sum_{{\mathbf n},\boldsymbol{\tau}}\prod_{i=1}^L
P(n_i,\tau_i)\,\delta\bigg(\!\sum_{i=1}^L n_i - N\bigg).
\end{equation}

As with the Marokovian ZRP, the factorized stationary distribution
provides the means for an analytic treatment of the model. In the
thermodynamic limit, the single-box probabilities in the steady
state $P(n,\tau)$ are equal to those of a single box with a
``mean-field'' incoming current $J$ which is generated by all other
sites. The master-equation for the single site-box probability is
\begin{equation}\label{eq:mfmaster}
\fl \frac{dP(n,\tau)}{dt} =  J P(n\!-\!1) \delta_{\tau,0} + c
P(n,\tau\!-\!1) + u(n\!+\!1,\tau)P(n\!+\!1,\tau) -
P(n,\tau)[J+c+u(n,\tau)],
\end{equation}
where the marginal occupation distribution is defined by $P(n)
\equiv \sum_{\tau} P(n,\tau)$. The first term on the RHS corresponds
to the box reaching the state $(n,\tau=0)$ by a particle entering a
box with $n-1$ particles, the second to an advance of the clock into
state $\tau$, the third to a particle leaving a box with $n+1$
particles, and the last to these three processes occurring when the
box is in state $(n,\tau)$. This equation is also valid for $n=0$ or
$\tau=0$ if one defines $P(-1,\tau) = P(n,-1) = 0$ (and, as stated
above, $u(0,\tau) = 0$ must also hold). Once \eqn
(\ref{eq:mfmaster}) is solved for a given MF current $J$, the
current and the probability distribution are obtained by the
self-consistency requirement
\begin{equation}
J = \sum_{n,\tau}u(n,\tau) P(n,\tau).
\end{equation}

In the steady state, $dP(n,\tau)/dt = 0$ and the master equation
(\ref{eq:mfmaster}) yields
\begin{equation}\label{eq:mfstationarymastereq}
P(n,\tau)[J + c + u(n,\tau)] = J P(n\!-\!1) \delta_{\tau,0} + c
P(n,\tau\!-\!1) + u(n\!+\!1,\tau)P(n\!+\!1,\tau).
\end{equation}
Summing over all values of $\tau$, the terms containing $c$ drop out
telescopically, and one is left with the recursion relation
\begin{equation}\label{eq:simplerec}
{\bar{u}}(n)P(n) = J P(n\!-\!1)
\end{equation}
Here $\bar{u}(n)$ is the mean hopping rate out of a site with $n$
particles
\begin{equation}\label{eq:ubar}
\bar{u}(n) \equiv \frac{\sum_{\tau}
P(n,\tau)\,u(n,\tau)}{\sum_{\tau} P(n,\tau)}.
\end{equation}
Equation (\ref{eq:simplerec}) expresses the balance between the
probabilities to jump into and out of a box with $n$ particles.
Iterating relations (\ref{eq:simplerec}) yields, as in the Markovian
ZRP, the steady-state occupation probability
\begin{equation}\label{eq:singlesiteprob}
P(n) = P(0)J^n\bar{f}(n),
\end{equation}
where the single-site weights are given by
\begin{equation}\label{eq:singlesiteweight}
\bar{f}(n) = \prod_{k=1}^{n}\bar{u}(k)^{-1},
\end{equation}
and $P(0)^{-1}=1+\sum_{n=1}^{\infty}J^n \bar{f}(n)$ ensures the
proper normalization of $P(n)$. The marginal distribution
(\ref{eq:singlesiteprob}) and (\ref{eq:singlesiteweight}) is the
same distribution one obtains for a Markovian ZRP but with the jump
rates $u(n)$ replaced by the effective rate $\bar{u}(n)$
\cite{evanszrpreview,andjel1982}.

Since the stationary distribution of our model has a similar form to
that of a Markovian ZRP, the analysis of condensation in the model
may also proceeds in a similar fashion. We therefore briefly review
how condensation takes place in the Markovian ZRP
\cite{evanszrpreview}. The occurrence of condensation in the
Markovian ZRP is determined by the asymptotic behavior of the jump
rates $u(n)$ for large $n$. Two types of condensation may be
distinguished: strong condensation, which occurs when the hopping
rates tend to zero for large $n$, and weak condensation, which may
occur when the hopping rates decrease to a constant value. In
condensation of the strong type all particles accumulate in one box
and the current vanishes in the thermodynamic limit. This
condensation occurs at all densities (i.e., the critical density for
condensation is $\rho_c = 0$). Weak condensation takes place only
when the rates decrease to a constant more slowly than $1+2/n$. In
particular, when the rates have the form
\begin{equation}\label{eq:jumprate}
u(n) = \gamma \bigl(1 + b/n^\sigma + o(1/n^\sigma)\bigr)
\end{equation}
for large $n$, condensation occurs above some critical density
provided that $\sigma < 1$ or $\sigma = 1$ and $b>2$. The critical
density $\rho_c$ is non-universal, i.e., it depends on the exact
form of the rates $u(n)$. Importantly, the parameter $\gamma$ only
sets the time scale for the process and has no affect on the
stationary distribution and the condensation transition.

In the weak condensation scenario, a single site (the condensate),
chosen spontaneously at random, accommodates $L(\rho-\rho_c)$
particles, while the density at all other sites remains $\rho_c$.
The condensation transition is thus manifest in the occupation
probability of a single site, $P(n)$. For the marginal case of
$\sigma = 1$, the probability to find $n$ particles in a given site
decays exponentially as $P(n) \sim n^{-b}e^{-n/\xi}$ for densities
below the critical density, where $\xi(\rho)$ diverges as $\rho_c$
is approached. At the critical density, the occupation probability
has a power law tail $P(n) \sim n^{-b}$. Above the critical density,
the occupation of all background sites remains power-law
distributed, while the occupation of the condensate is narrowly
distributed around $L(\rho-\rho_c)$
\cite{EvansMajumdarZia2006CanonicalZRP}.

\subsection{Condensation in an ``on-off'' model}\label{sec:MFonoff}
As the effective jump rates $\bar{u}(n)$ play the role of $u(n)$ in
the non-Markovian ZRP, it is their asymptotic behavior for
$n\to\infty$ which determines condensation in the model. The
remainder of this section concentrates on the determination of this
asymptotic behavior. We begin by discussing a simple choice of jump
rates --- an ``on-off'' model which will now be introduced
--- before turning to an analysis of more general jump rates.

We start the discussion by considering jump rates of the form
\begin{equation}\label{eq:onoffrates}
u(n,\tau) = \left\{
\begin{array}{ll} 0 & \tau =0 \quad\mytext{(``off'' state)}\\
u(n) & \tau \geq 1 \quad\mytext{(``on'' state).}
\end{array} \right.
\end{equation}
In this case, every time a particle hops into a box, that box is
turned ``off''. When the box is in this off state no particle can
leave it. After an exponentially distributed random time (with
parameter $c$) the box is turned back ``on'', and particles can once
more jump out of it with a rate $u(n)$. The model with these special
rates will be called the \emph{on-off model}.

In the on-off model, the dynamics depends only on whether $\tau = 0$
or $\tau \geq 1$, and thus the clock has effectively only two
states. Correspondingly, the state of a box can be characterized by
$P_\mysubtext{off}(n) \equiv P(n,\tau=0)$ and $P_\mysubtext{on}(n) =
\sum_{\tau\geq 1}P(n,\tau)$. The stationary master equation
(\ref{eq:mfstationarymastereq}) is then given by
\begin{eqnarray}
P_\mysubtext{off}(n)[J + c] &=& J P(n\!-\!1) \label{eq:onoffmastereq1} \\
P_\mysubtext{on}(n)[J + u(n)] &=& c P_\mysubtext{off}(n) +
P_\mysubtext{on}(n\!+\!1)u(n\!+\!1) \label{eq:onoffmastereq2} .
\end{eqnarray}
The solution of these equations is made simple, compared with a
general non-Markovian ZRP, because the term $p(n+1)u(n+1,0)$ which
should appear in the RHS of (\ref{eq:onoffmastereq1}) (see \eqn
(\ref{eq:mfstationarymastereq})) vanishes.

To solve these equations we first note that by summing \eqn
(\ref{eq:onoffmastereq1}) over all values of $n$ we find that the
probability to find a site in the off state is
\begin{equation}\label{eq:poff}
P_{\mysubtext{off}} \equiv \sum_n P_{\mysubtext{off}}(n) =
\frac{J}{c\!+\!J}.
\end{equation}
Next, an expression for $P_\mysubtext{off}(n)$ is found from \eqn
(\ref{eq:onoffmastereq1}) together with (\ref{eq:simplerec}) and
(\ref{eq:poff}),
\begin{equation}\label{eq:onofffullprob1}
P_{\mysubtext{off}}(n) = \frac{\bar{u}(n)P(n)}{J+c} =
\frac{P_{\mysubtext{off}}}{J}\bar{u}(n)P(n).
\end{equation}
A similar  expression for $P_\mysubtext{on}(n)$ is found by
substituting the rates $u(n,\tau)$ (which are of the form
(\ref{eq:onoffrates})) into \eqn (\ref{eq:ubar}), yielding
\begin{equation}\label{eq:onofffullprob2}
P_{\mysubtext{on}}(n) = \frac{\bar{u}(n)}{u(n)}P(n).
\end{equation}

The effective jump rates $\bar{u}(n)$ can now be obtained from
(\ref{eq:onofffullprob1}) and (\ref{eq:onofffullprob2}) using
$P_\mysubtext{off}(n) + P_\mysubtext{on}(n) = P(n)$, and are given
by
\begin{equation}\label{eq:onoffubar}
\frac{1}{\bar{u}(n)} = \frac{P_{\mysubtext{off}}}{J}+\frac{1}{u(n)}.
\end{equation}
This equation states that the mean time between hops from a site
with $n$ particles is equal to the mean time this site is in an
``off'' state plus the time it takes a particle to hop out once the
system is already ``on''.

Using \eqns (\ref{eq:singlesiteprob}), (\ref{eq:singlesiteweight}),
(\ref{eq:poff}) and (\ref{eq:onoffubar}), it is now possible to
obtain $P(n)$ for any $J$, and subsequently the entire probability
distribution is found via (\ref{eq:onofffullprob1}) and
(\ref{eq:onofffullprob2}). Note that $P(n)$ which is found this way
depends on $J$ both directly, as seen in \eqn
(\ref{eq:singlesiteprob}), and indirectly through the effective
rates (\ref{eq:onoffubar}). To finish the calculation, one must find
the dependence of the current $J$ on the density $\rho$. This can in
principle be achieved by inverting the relation $\rho(J) = \sum_n n
P(n)$. The effective hopping rates (\ref{eq:onoffubar}) are thus a
function of the density.

To determine whether or not condensation may occur in the model,
only the asymptotic form of $\bar{u}(n)$ is needed. By examining
\eqn (\ref{eq:onoffubar}) it is seen that $\bar{u}(n)$ decreases to
zero when $n\to \infty$ if and only if $u(n)$ decreases to zero, and
similarly $\bar{u}(n)$ decreases to a constant if and only if $u(n)$
decreases to a constant. Therefore, strong condensation is not
affected by the clock-dependent dynamics. To study weak
condensation, assume jump rates of the asymptotic form
(\ref{eq:jumprate}) with $\gamma = 1$ (as explained above, $\gamma$
sets the time scale of the process, and can be set to 1 without loss
of generality). From \eqns (\ref{eq:poff}) and (\ref{eq:onoffubar})
we find, to leading order in $1/n$
\begin{equation}\label{eq:onoffubarfinal}
\bar{u}(n) =
\frac{c+J}{c+J+1}\bigg(1+\frac{b_{\mysubtext{eff}}}{n^\sigma} +
o\Bigl(\frac{1}{n^{\sigma}}\Bigr)\bigg),
\end{equation}
which is again of the form (\ref{eq:jumprate}) but with an effective
hopping parameter
\begin{equation}\label{eq:mfbeff}
b_{\mysubtext{eff}}=\frac{c+J}{c+J+1}\, b \, < \, b.
\end{equation}
If $\sigma < 1$ condensation occurs in the on-off model as it does
in the Markovian ZRP. In the commonly encountered case of $\sigma =
1$, however, condensation only occurs when $b_\mysubtext{eff}>2$.
The critical current at the condensation transition is given in this
case by $J_c = \lim_{n\to\infty}\bar{u}(n)$ \cite{evanszrpreview},
which yields, according to (\ref{eq:onoffubarfinal}), $J_c =
{(c+J_c)}/{(c+J_c+1)}$, or
\begin{equation}\label{eq:onoffJc1b}
J_c = \frac{c}{2}\bigg(\sqrt{1+\frac{4}{c}}-1\bigg).
\end{equation}
This allows us to write
\begin{equation}\label{eq:mfbeff1}
b_\mysubtext{eff} = J_c\, b.
\end{equation}
Therefore, condensation takes place when the hopping parameter
satisfies $b
> \frac{4}{c}\left(\sqrt{1+4/c}-1\right)^{-1}$. The critical value of $b$ is larger than
2, in contrast with the Markovian case for which the critical value
for condensation is $b=2$.

\subsection{Condensation in MF models with more general rates}
In the previous section we have seen that in the case of jump rates
with an asymptotic form
\begin{equation}\label{eq:jumprate1}
u(n) = 1 + b/n + ...,
\end{equation}
the on-off dynamics leads to an effective value of $b$, and thus it
may affect the occurrence of the condensation transition. We now
demonstrate that this holds also when the clock dependence is more
general than the on-off case, and we show how $b_\mysubtext{eff}$
may be calculated. To this end we consider rates of the form
\begin{equation}\label{eq:jumprate2}
u(n,\tau) = u(n)v(\tau) = \Bigl(1+\frac{b}{n}+\ldots\Bigr)v(\tau)
\end{equation}
where $u(n)$ has been taken to be of the form (\ref{eq:jumprate1}).

As mentioned above, the stationary master equation
(\ref{eq:mfstationarymastereq}) is harder to analyze when the rates
are not of the on-off type, because $P(n,\tau)$ depends in
(\ref{eq:mfstationarymastereq}) on $P(n,\tau+1)$. However, since
only the large $n$ asymptotics of $\bar{u}(n)$ and $P(n,\tau)$ at
criticality affect condensation, it is possible to make progress by
restricting the discussion to these quantities. We therefore assume
that $J=J_c$ and make the following ansatz:
\begin{eqnarray}\label{eq:mfgenansatz}
\bar{u}(n) = J_c\bigg[1 + \frac{b_\mysubtext{eff}}{n} +
o\Bigl(\frac{1}{n}\Bigr) \bigg]  \nonumber \\
P(n,\tau) = A
n^{-b_\mysubtext{eff}}\,\alpha(\tau)\bigg[1+\frac{d(\tau)}{n}+
o\Bigl(\frac{1}{n}\Bigr) \bigg]
\\
P(n) = A
n^{-b_\mysubtext{eff}}\bigg[1+\frac{d}{n}+o\Bigl(\frac{1}{n}\Bigr)\bigg].
\nonumber
\end{eqnarray}
This ansatz is motivated by the solution of the on-off model
(compare with \eqns (\ref{eq:onofffullprob1}) and
(\ref{eq:onoffubarfinal})). The constant $A$ is a normalization
constant, and from the definition $P(n) = \sum_\tau P(n,\tau)$ it is
seen that $\sum_\tau \alpha(\tau) = 1$ and $\sum_\tau
\alpha(\tau)d(\tau) = d$ must hold.

Substituting the ansatz (\ref{eq:mfgenansatz}) in the stationary
master equation (\ref{eq:mfstationarymastereq}) and equating terms
order by order in $1/n$ yields to order O(1)
\begin{equation}\label{eq:mfgenalpha}
\alpha(\tau) = \frac{J_c}{J_c+c}\biggl(\frac{c}{J_c+c}\biggr)^\tau
\end{equation}
and to order $O(1/n)$
\begin{equation}\label{eq:mfgend}
d(\tau) = d - \biggl[\frac{1}{J_c + c}\sum_{\tau' = 0}^\tau v(\tau')
- 1\biggr]b_\mysubtext{eff}.
\end{equation}
The current and $b_\mysubtext{eff}$ can now be found by substituting
(\ref{eq:mfgenansatz}) in the definition of $\bar{u}(n)$ (\eqn
(\ref{eq:ubar})) and equating once again order by order in $1/n$. To
order $O(1)$, an equation for the critical current is obtained
\begin{equation}\label{eq:mfgenJ}
J_c = \sum_{\tau=0}^\infty \alpha(\tau)v(\tau) = \frac{J_c}{J_c + c}
\sum_{\tau=0}^\infty \biggl(\frac{c}{J_c + c}\biggr)^\tau v(\tau),
\end{equation}
where (\ref{eq:mfgenalpha}) was used in the last equality. To order
$O(1/n)$, using (\ref{eq:mfgenalpha}) and (\ref{eq:mfgend}),
$b_\mysubtext{eff}$ is found to satisfy
\begin{eqnarray}\label{eq:mfgenbeff}
b_\mysubtext{eff} &=& J_c(J_c+c)\biggl[\sum_{\tau=0}^\infty
\sum_{\tau'=0}^\tau \alpha(\tau) v(\tau)  v(\tau') \biggr]^{-1}
\cdot b  \nonumber
\\
&=&  (J_c+c)^2\biggl[\sum_{\tau=0}^\infty \sum_{\tau'=0}^\tau
\biggl(\frac{c}{J_c + c}\biggr)^\tau v(\tau) v(\tau') \biggr]^{-1}
\cdot b
\end{eqnarray}

The calculation outlined above is valid as long as the series in
(\ref{eq:mfgenJ}) and (\ref{eq:mfgenbeff}) converge. The exponential
form of $\alpha(\tau)$ in \eqn (\ref{eq:mfgenalpha}) implies that
convergence is guaranteed if $v(\tau)$ decays or grows slower than
exponentially. In particular, this implies that the results are
correct if $v(\tau)$ tends to a finite (non-zero) constant for large
$\tau$. Note that if $v(\tau)$ decays to zero fast enough, although
the series converge any system of a finite size will eventually be
frozen in an absorbing state in which all $\tau$'s tend to infinity
and no particles jump.

\begin{figure}
\center
  \includegraphics[width=0.5\textwidth]{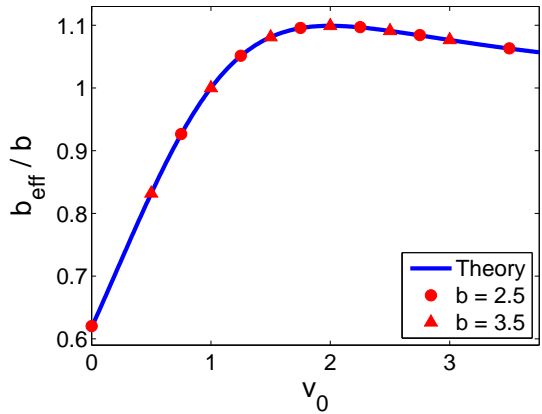}
  \caption{ \label{fig:MFgen2State}{The mean-field value of
  $b_\mysubtext{eff}/b$ as a function of $v_0$ for the
  rates (\ref{eq:weakstrongrates}). The line corresponds to the
  prediction (\ref{eq:mfgenbeff}), while the symbols were obtained
  by numerically integrating the mean-field master equation
  (\ref{eq:mfmaster}) with $J=J_c$ of (\ref{eq:mfgenJ}).
  Note that $b_\mysubtext{eff}$ may be either smaller or larger than $b$,
  depending on $v_0$.
  }}
\end{figure}

Equation (\ref{eq:mfgenbeff}) implies that, as found in the
particular case of the on-off model, the condensation behavior
depends on the memory effects induced by the clocks. Note, however,
that unlike the on-off case, $b_\mysubtext{eff}$ is not necessarily
smaller than $b$. For instance, consider rates of the form
(\ref{eq:jumprate2}) with
\begin{equation}\label{eq:weakstrongrates}
v(\tau) = \left\{
\begin{array}{ll}
v_0 & \tau = 0 \\
1 & \tau \geq 1
\end{array}\right..
\end{equation}
For $v_0 = 0$ these rates reduce to the on-off model, while $v_0 =
1$ is the Markovian ZRP. For arbitrary $v_0$, \eqn (\ref{eq:mfgenJ})
yields $J_c = (v_0-c+\sqrt{(v_0-c)^2+4c})/2$, and \eqn
(\ref{eq:mfgenbeff}) yields $b_\mysubtext{eff} = b\,(c+v_0
J_c)/(c+v_0 J_c + J_c - v_0)$. This result, which is plotted in \fig
\ref{fig:MFgen2State}, demonstrates that for different values of
$v_0$ and $c$, the effective hopping parameter $b_\mysubtext{eff}$
might be larger or smaller than the ``bare'' value $b$.

\section{The non-Markovian ZRP with nearest-neighbor
dynamics}\label{sec:Ring}

The results of the previous section demonstrate that temporal
correlations in the dynamics of a mean-field ZRP affect the
condensation transition. In this section we examine whether this
mean-field picture persists also when the dynamics allows only
nearest-neighbor hopping, and whether new effects appear in the
latter case.

In the on-off model with nearest-neighbor hopping dynamics, the
stationary distribution does not factorize and the stationary
solution of the Master equation is not known. We therefore study the
model using numerical Monte-Carlo simulations. From these
simulations we find that condensation does indeed seem to be
controlled by an effective hopping parameter $b_\mysubtext{eff}$,
albeit with a value which differs from the MF prediction. We also
find that asymmetric jump rates may cause the condensate to drift
with a finite velocity.

In this section we concentrate solely on the on-off model with jump
rates of the form (\ref{eq:onoffrates}) and (\ref{eq:jumprate1}),
unless explicitly stated otherwise.

\subsection{On-off model with symmetric nearest-neighbor hopping}
We begin the discussion of a ring with nearest-neighbor hopping
dynamics by considering an on-off model with symmetric hopping,
i.e., with $p = 1/2$. Note that, unlike the Markovian ZRP with
symmetric hopping, which satisfies detailed balance and hence is an
equilibrium model, the non-Markovian ZRP does not satisfy detailed
balance even when it is symmetric. To understand why, note that
there are allowed dynamical moves whose reverse cannot occur (such
as an advance of a clock, or a jump of a particle simultaneously
with resetting the clock of the target site to zero). As these moves
have a non-zero probability to occur in the steady state, stationary
probability currents must exist.

\begin{figure}[!t]
  \center
  \subfloat[]{\label{fig:symmetricpofon}\includegraphics[width=0.49\textwidth]{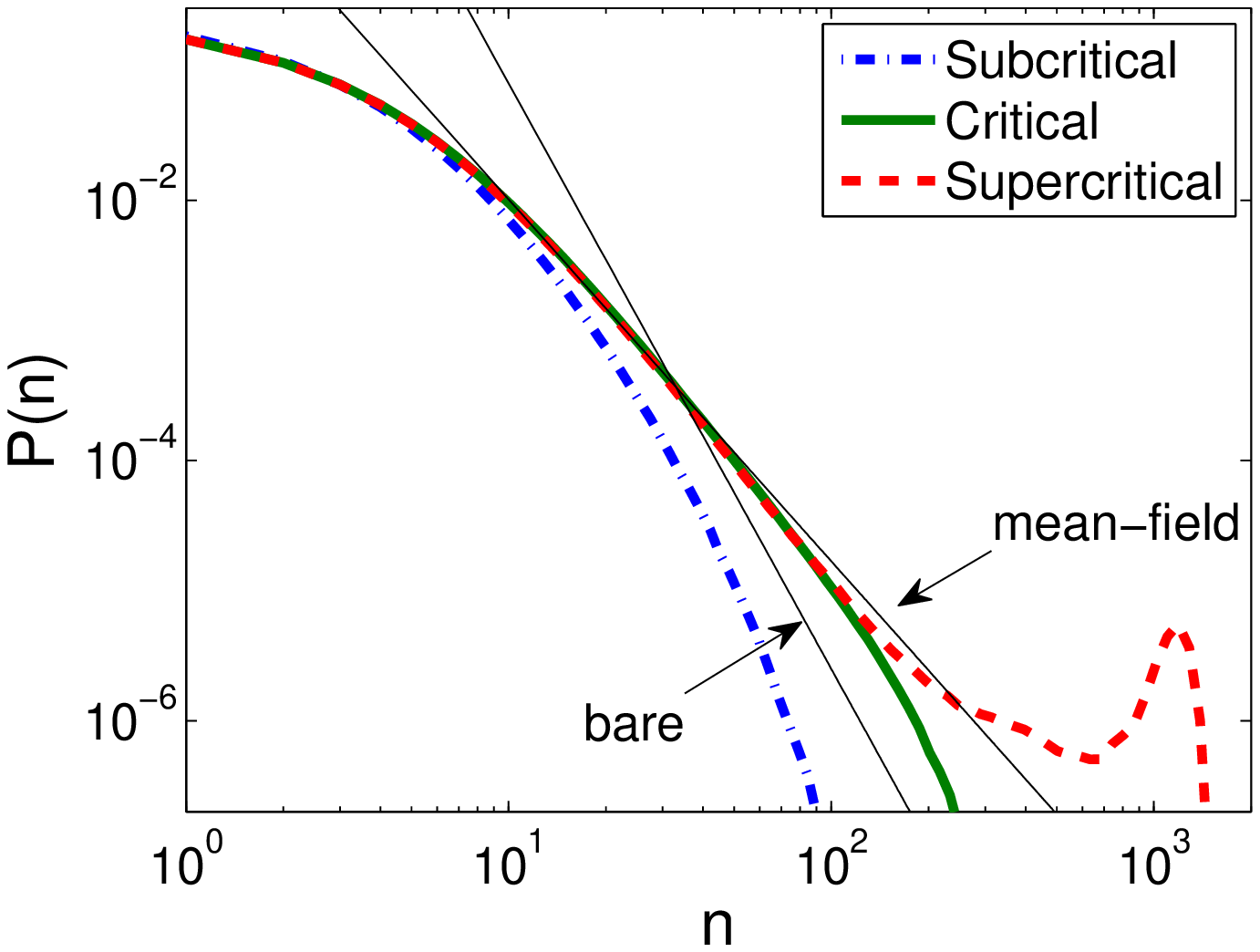}}
  \subfloat[]{\label{fig:symmetriclattice}\includegraphics[width=0.49\textwidth]{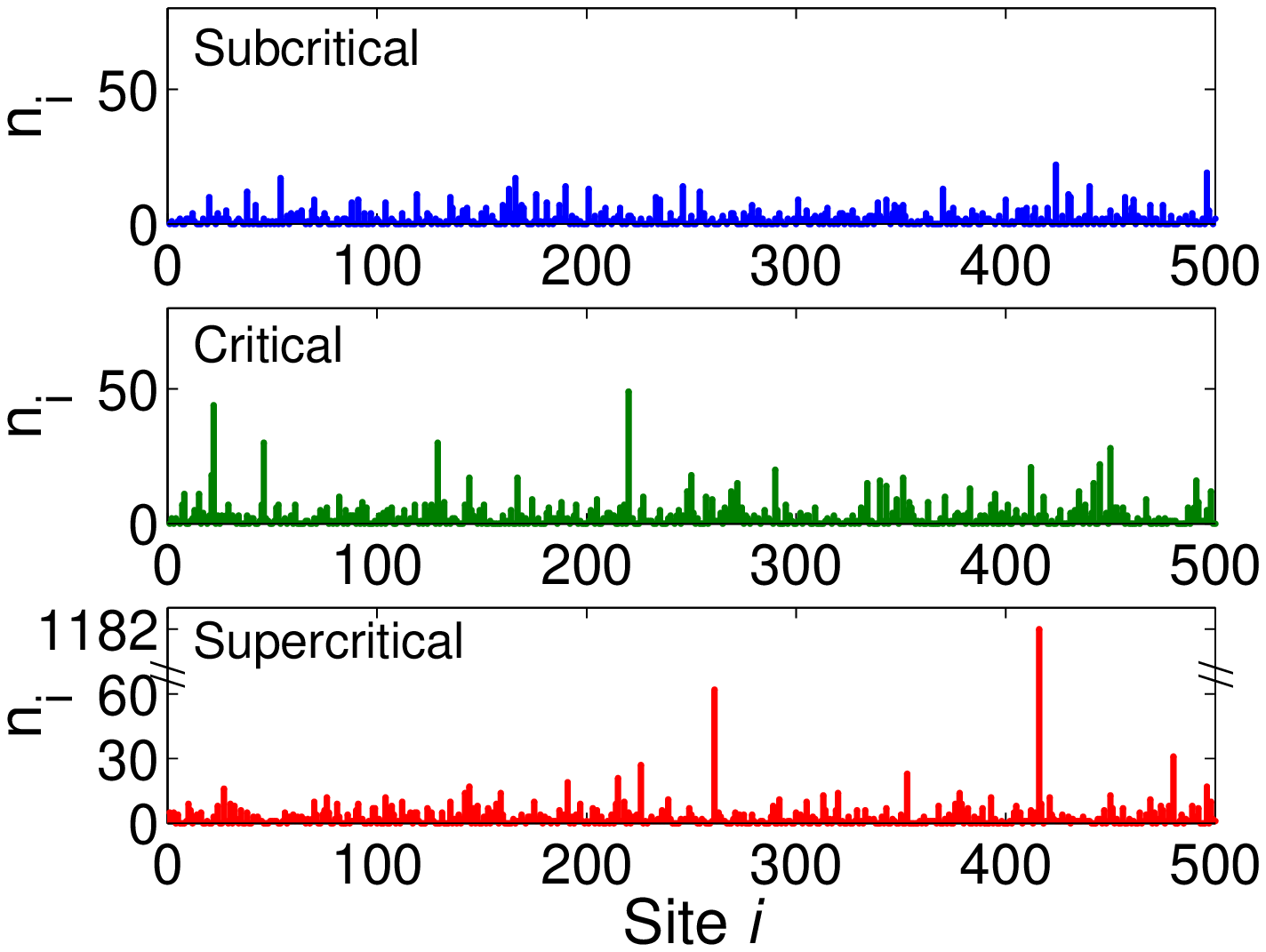}}
  \caption{{(a) The occupation probability $P(n)$ of a single
  site in a ring with symmetric nearest-neighbor on-off dynamics for several
  densities. (b) Typical snapshots of the lattice
  for several densities. Results were obtained by Monte-Carlo
  simulations with $b=4.5$, $c=1$ and $L=500$ sites,
  for a subcritical
  density ($\rho = 2$), a supercritical density ($\rho = 5$) and at
  the critical region ($\rho = 2.75$). The two thin straight line in (a)
  correspond to the power laws $P(n) \sim n^{-b}$ and $P(n) \sim n^{-b_\mysubtext{eff}}$ of
  the bare value $b = 4.5$ and the mean-field effective value
  $b_\mysubtext{eff} \approx 2.78$ of \eqn (\ref{eq:mfbeff1}).
  The power-law exponent for symmetric dynamics is seen to lie between
  these two values. Note that in the supercritical case,
  the y-axis of figure (b) is broken in order to show both the condensate and the
  disordered background.
  }}
  \label{fig:ringsymmetric}
\end{figure}

Monte-Carlo simulations of the on-off model with nearest neighbor
hopping were carried out on a ring of size $L=500$ boxes with
different particle densities and $b=4.5$. The system was initialized
to a state in which all particles were located at the first site and
all sites were ``on'', and the dynamics was run for a time of
$t_\mysubtext{equil} = 2.5\cdot10^{7}$ time units to allow the
system to reach a steady state. After this equilibration time, the
state of the system was recorded every $t_\mysubtext{sampling} =
5000$ time units. The measured single site occupation probability
$P(n)$ and typical snapshots of the lattice for different values of
$\rho$, presented in \fig \ref{fig:ringsymmetric}, show a
qualitative resemblance to those of a Markovian ZRP. At small
densities when the system is in the fluid phase, the single site
occupation probability has an exponential tail, while at high
densities this probability develops a peak which corresponds to the
condensate. The transition takes place at a critical density (which
is found to be $\rho_c \approx 2.75$) at which the occupation
probability decays as a power law of the form $P(n) \sim
n^{-b_\mysubtext{eff}}$. In finite systems this power law has an
exponential cutoff due to finite size effects.

Measuring the effective hopping parameter $b_\mysubtext{eff}$
numerically is a difficult task because it depends on the tail of
the probability distribution which is strongly distorted by finite
size effect. However, simulation results indicate that
$b_\mysubtext{eff}$ for symmetric hopping is larger than the MF
value (\ref{eq:mfbeff1}) and smaller than the ``bare'' value $b$
(see Figure \ref{fig:ringsymmetric}).

The simulation results indicate that the conclusions which were
found for mean-field dynamics are qualitatively correct for
symmetric nearest-neighbor dynamics.

\subsection{On-off model with asymmetric nearest-neighbor hopping}

Simulations of asymmetric nearest-neighbor dynamics (i.e., with
$p<1/2$) indicate that, as with the symmetric case, condensation is
controlled by an effective hopping parameter. However, a new effect
is found in simulations of asymmetric hopping: the condensate drifts
with a finite velocity. Two different drift regimes are observed in
simulations of finite systems: a ``strong drift'' regime in which
the condensate is in a continual motion, and a ``weak drift'' regime
in which the condensate stays for some (random) time in each site
before jumping to the next. In what follows we begin by discussing
the case of totally asymmetric hopping dynamics, (i.e., with
asymmetry parameter $p=0$), where we examine the strong drift and
the weak drift regimes separately. We then discuss more general
asymmetric dynamics, including partially asymmetric hopping and
asymmetric dynamics on higher dimensional lattices.

\subsubsection{Totally asymmetric hopping: strong drift regime}
\begin{figure}
\center
  \includegraphics[width=0.5\textwidth]{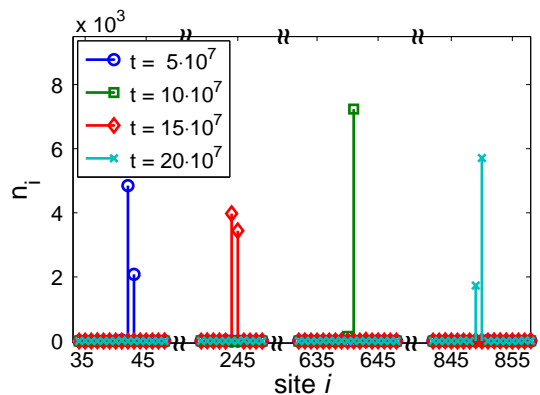}
  \caption{ \label{fig:ringcondensatesnapshot}{Snapshots of the on-off model with
 totally-asymmetric
  nearest-neighbor hopping on a ring, showing the occupation numbers $n_i$ at and in the
vicinity of the condensate at four points in time.
  Here, $L=1000$ sites, $\rho = 10$,
  $b = 5.5$, and $c=1$.
  The condensate occupies two sites
  and drifts with a constant mean velocity.
  }}
\end{figure}

Monte-Carlo simulations of a ring with totally asymmetric
nearest-neighbor on-off hopping dynamics were carried out for
different values of $c$. The results of these simulations show that
for small values of $c$ the condensate drifts continuously in what
we term a strong drift regime. In this regime, the condensate
typically occupies two adjacent boxes $i$ and $i+1$, in contrast to
previously known condensation phenomena. In addition, The location
of these two boxes advances with time. This is demonstrated in Fig.\
\ref{fig:ringcondensatesnapshot}, where we present snapshots of the
lattice taken at different times as obtained from a simulation with
$L=1000$ boxes.

An inspection of the microscopic dynamics shows that the drift of
the condensate takes place via a ``slinky'' motion in which the
second condensate site, $i+1$, accumulated particles at the expense
of the first condensate site, $i$. This slinky motion results from
the fact that site $i+1$ is turned off more often than other sites.
In other words, the effective hopping rates out of a site are no
longer homogeneous in space, but rather they depend on the distance
of the site from the condensate, and in particular, the mean current
out of the condensate is larger than the mean current out of the
next site: $\langle \bar{u}_i(n_i)\rangle
> \langle \bar{u}_{i+1}(n_{i+1})\rangle$. Thus particles
accumulate on site $i+1$ until site $i$ is no longer macroscopically
occupied, giving the clock at $i+1$ the chance to reach the on state
for durations of time sufficiently long to allow particles to
escape. Then particles start to hop from site $i+1$ to site $i+2$ in
the same fashion and the slinky motion continues. This mechanism for
condensate motion was recently found in other models, and will be
analyzed in more detail elsewhere \cite{ToyModel}.

This slinky motion mechanism suggests that the drift velocity
$v_\mysubtext{drift}$ is inversely proportional to the number of
particles in the condensate $N_\mysubtext{cond}$, i.e.,
\begin{equation}\label{eq:condensatevel}
v_\mysubtext{drift}^{-1}\sim N_\mysubtext{cond} = N-N_\mysubtext{bg}
= L(\rho-\rho_\mysubtext{bg}),
\end{equation}
where $N_\mysubtext{bg}\equiv N-N_\mysubtext{cond}$ and
$\rho_\mysubtext{bg}\equiv N_\mysubtext{bg}/(L-2)$ are respectively
the mean number and the density of particles in the background
fluid, i.e., in all sites but the condensate sites. In the
thermodynamic limit, the velocity of the condensate vanishes. It
should be noted that this drift motion of the condensate is
different from the relocation of the condensate which occurs in
Markovian ZRPs. In the Markovian case, a condensate on any finite
system can melt and reappear at some other randomly chosen site of
the lattice. This relocation of the condensate happens on a
characteristic time which scales with the system size to a power
larger than 2
\cite{grosskinskyetal2003zrpcondensation,godrecheluck2005condensate,beltramlandim2008,Landim2012AsymmetricZRPCondensate}.
A similar relocation of the condensate to a random distant site is
seen to occur also in the asymmetric on-off ZRP, superimposed on the
``slinky'' drift motion.

\begin{figure}
\center
    \subfloat[]{\label{fig:ringcondensatedrfit}\includegraphics[width=0.47\textwidth]{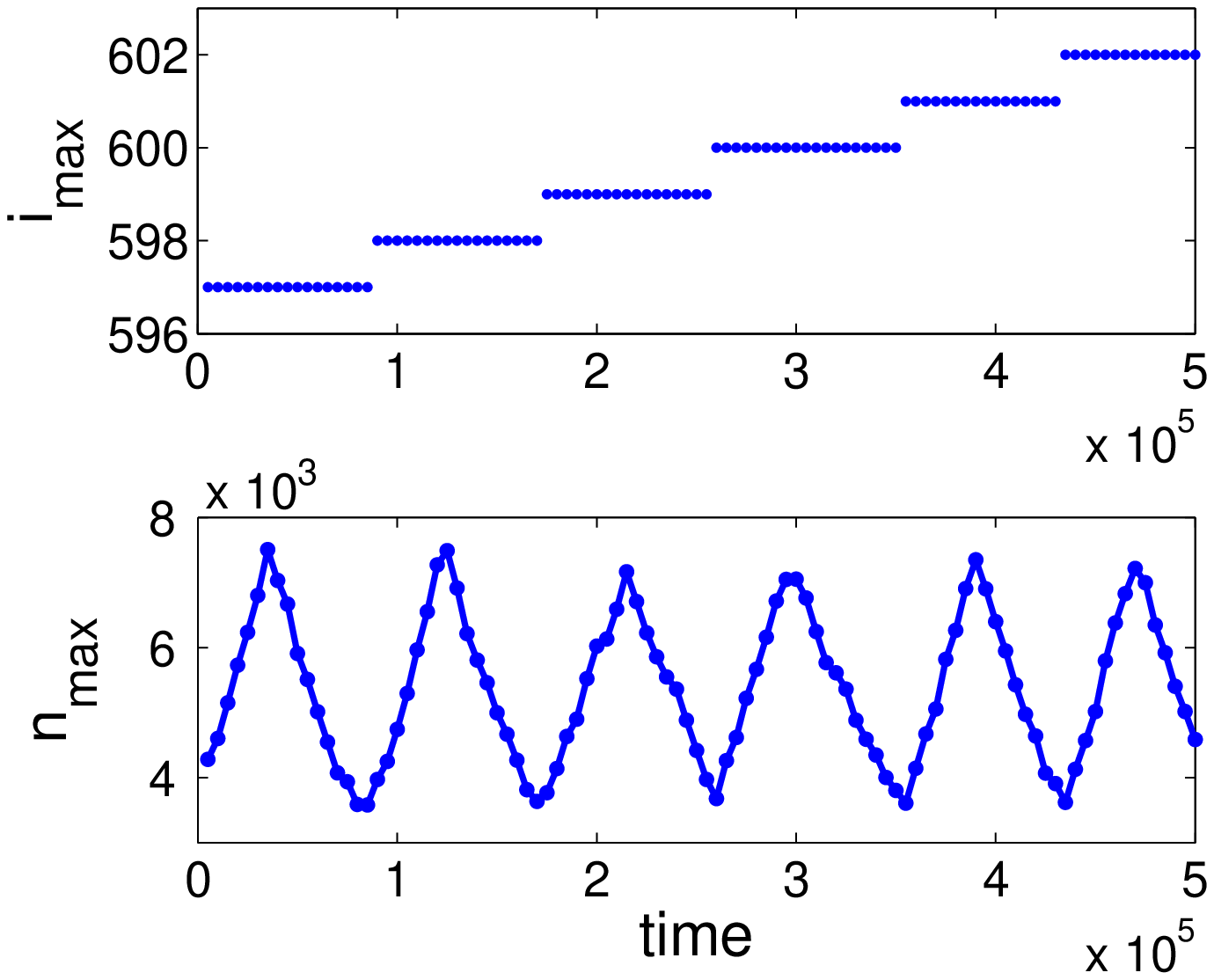}}
  \subfloat[]{\label{fig:condensatevelocity}\includegraphics[width=0.5\textwidth]{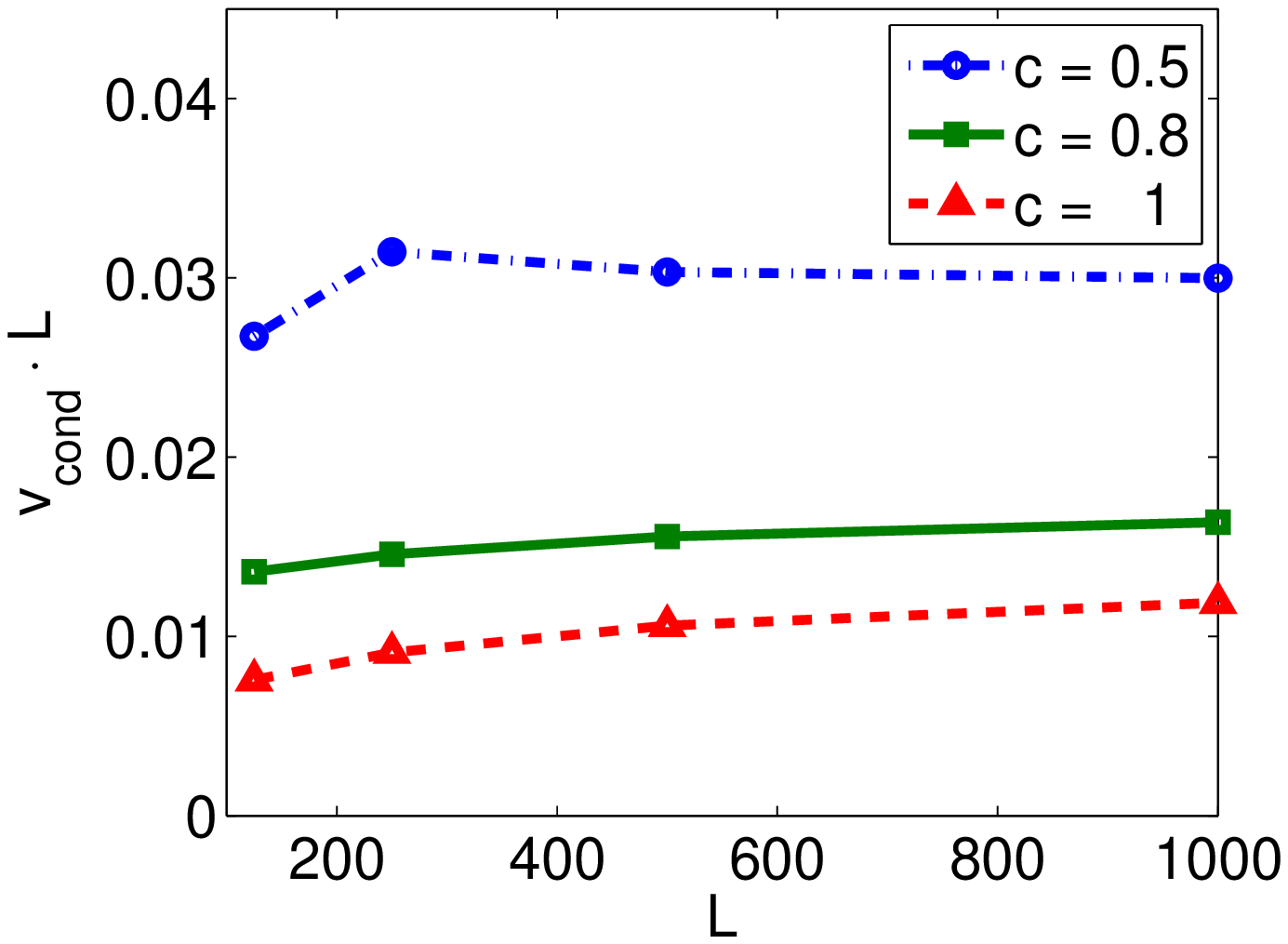}}
  \caption{{(a) The ``slinky'' motion of the condensate is seen
  in the position of the most occupied site
  $i_{\max}$ and its occupation number $n_{\max}$ as a
  function of time, on a ring of size $L=1000$ with $c=1$. (b) This slinky
  motion suggests that the velocity of the condensate scales as $L^{-1}$, as is
  demonstrated for different values of $c$. Graphs were obtained from
  simulation of  totally-asymmetric
  nearest-neighbor on-off dynamics with $\rho = 10$ and
  $b = 5.5$.
  }}
  \label{fig:condensatedrift}
\end{figure}

The snapshots presented in Fig.\ \ref{fig:ringcondensatesnapshot}
clearly demonstrate that the condensate occupies two adjacent sites
with varying relative occupation, consistent with the slinky motion
described above. In addition, the drift of the condensate is evident
in the figure. In order to demonstrate the slinky motion in more
detail, we present in Fig.\ \ref{fig:ringcondensatedrfit} a plot
showing the position of the most occupied site $i_\mysubtext{max}$
and its occupation number, $n_\mysubtext{max}$, as a function of
time. The occupation number $n_\mysubtext{max}$ oscillates in time
with approximately constant frequency. Typically it decreases
linearly until it reaches its minimal value, when
$i_\mysubtext{max}$ increases by 1 and $n_\mysubtext{max}$ starts
increasing. \fig \ref{fig:condensatevelocity} displays the scaling
of the condensate velocity with the system size $L$, which agrees
with the estimate of \eqn (\ref{eq:condensatevel}).

\begin{figure}
\center
  \subfloat[]{\includegraphics[width=0.49\textwidth]{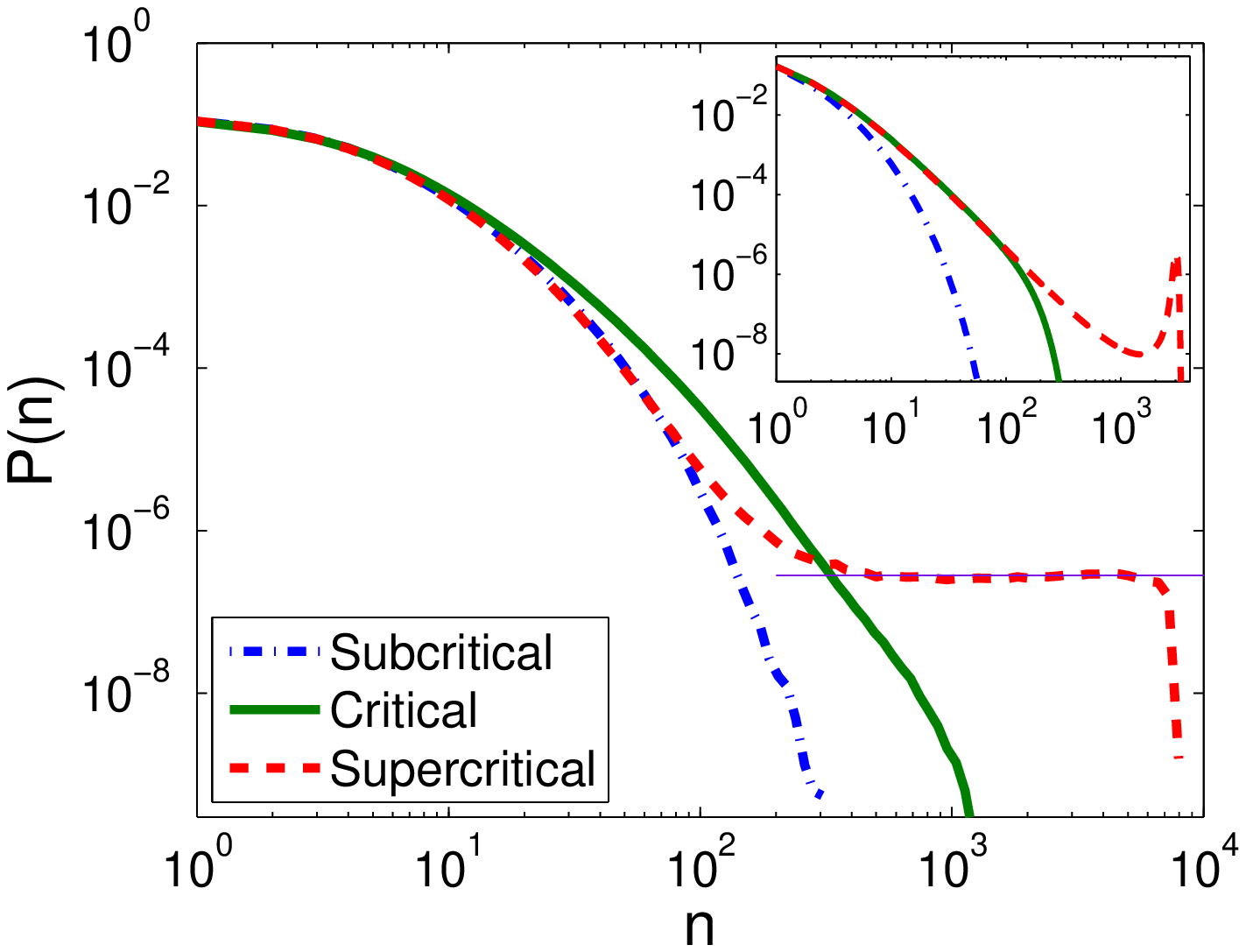}}
  \subfloat[]{\includegraphics[width=0.49\textwidth]{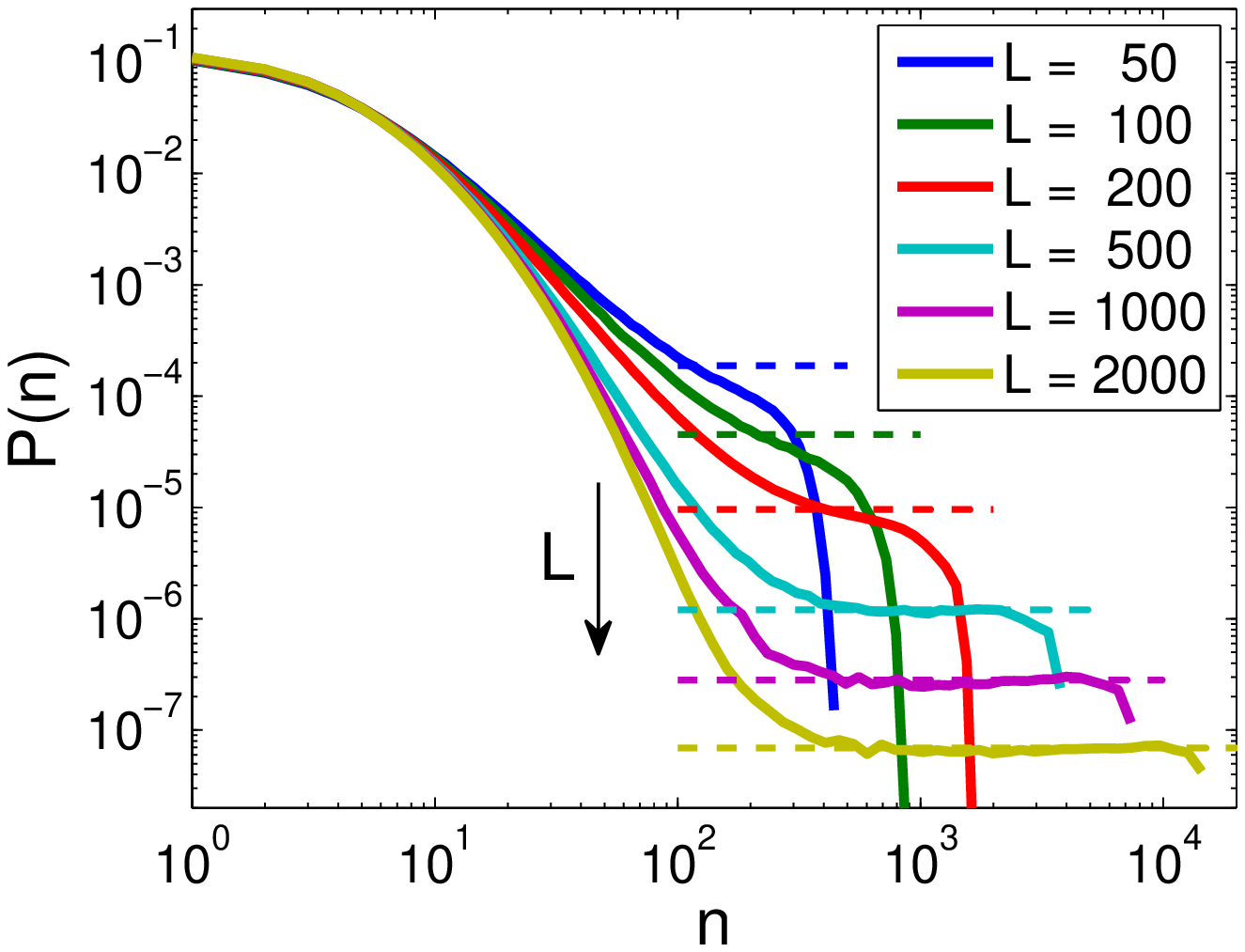}}
  \caption{{The occupation probability $P(n)$ of a single-site in a ring with
  totally-asymmetric nearest-neighbor hopping and different densities,
  as obtained from Monte-Carlo simulations. (a) $P(n)$
  at a subcritical density ($\rho = 3$), supercritical density
  ($\rho = 10$) and at the critical region ($\rho=4.1$).
  The horizontal
  line indicates $P_\mysubtext{plateau}$ of (\ref{eq:ringpcond}), where
  $\rho_\mysubtext{bg}$ was obtained from the simulation. Here, $L=1000$,
  $b=5.5$ and $c=1$.
  The inset shows for comparison a similar plot of $P(n)$ for a Markovian ZRP
  of $L=1000$ sites with $b=3$, in the subcritical ($\rho = 0.5$) critical
  ($\rho = 1$) and supercritical ($\rho = 4$) phases. (b) $P(n)$ for
  several system sizes ($L$ increases in the direction of the arrow). The dashed horizontal
  lines indicates $P_{\mysubtext{plateau}}$ of (\ref{eq:ringpcond}). Here
  $\rho=10$, $b=5.5$ and $c=1$.
  }}
  \label{fig:ringpofn}
\end{figure}

In Fig.\ \ref{fig:ringpofn} we present the single-site occupation
probability distribution $P(n)$ for various densities and for
various system sizes. At high densities the distribution exhibits a
plateau which reflects the particle distribution among the two sites
which constitute the condensate. This is in contrast with a
Markovian ZRP and the symmetric on-off model where the condensate is
supported by a single site, which results in a sharp peak in $P(n)$
(compare with the inset and with \fig \ref{fig:symmetricpofon}). The
value of $P(n)$ at the plateau in the non-Markovian case may be
estimated for $\rho$ above the critical density and large $L$ using
the slinky motion of the condensate. The probability that a given
site carries the condensate is $2/L$, and in such a site there is an
approximately uniform probability to find any occupation
$0<n<N_\mysubtext{cond}=L(\rho-\rho_\mysubtext{bg})$. Thus,
\begin{equation}\label{eq:ringpcond}
P_\mysubtext{plateau} \sim
\frac{2}{L}\,\frac{1}{L(\rho-\rho_\mysubtext{bg})}.
\end{equation}
This estimate is in good agreement with the plateau value in \fig
\ref{fig:ringpofn}. For small densities,  $P(n)$ decays
exponentially, indicating the absence of a condensate. For the
system size studied in this figure, the distribution at small values
of $n$ does not allow to extract a power law decay as expected for
the condensation transition. At density $\rho = 4.1$ there is a
range of $n$ for which $P(n)$ seems to follow a power law with
$b_\mysubtext{eff} \approx 4$. This value differs significantly from
the bare parameter $b = 5.5$, the expected value for Markovian ZRP.

\subsubsection{Totally asymmetric hopping: weak drift regime} For
larger values of the clock rate $c$, the motion of the condensate
looks qualitatively different from that in the strong drift regime:
the continuous slinky motion is replaced by an erratic slinky
motion, in which the condensate spends a long period of time in each
site before jumping to the next. We refer to this regime as the weak
drift regime. This difference is observed on finite systems. Whether
this type of motion persists for large $L$ remains an open question
at this point, as there are some indications that the thermodynamic
behavior might be similar to the strong drift motion in this limit.
In what follows we present the numerical evidence for the weak drift
regime, and provide details on the question of the thermodynamic
limit.

\begin{figure}
\center
  \subfloat[]{\label{fig:weakcondensatedrfit}\includegraphics[width=0.49\textwidth]{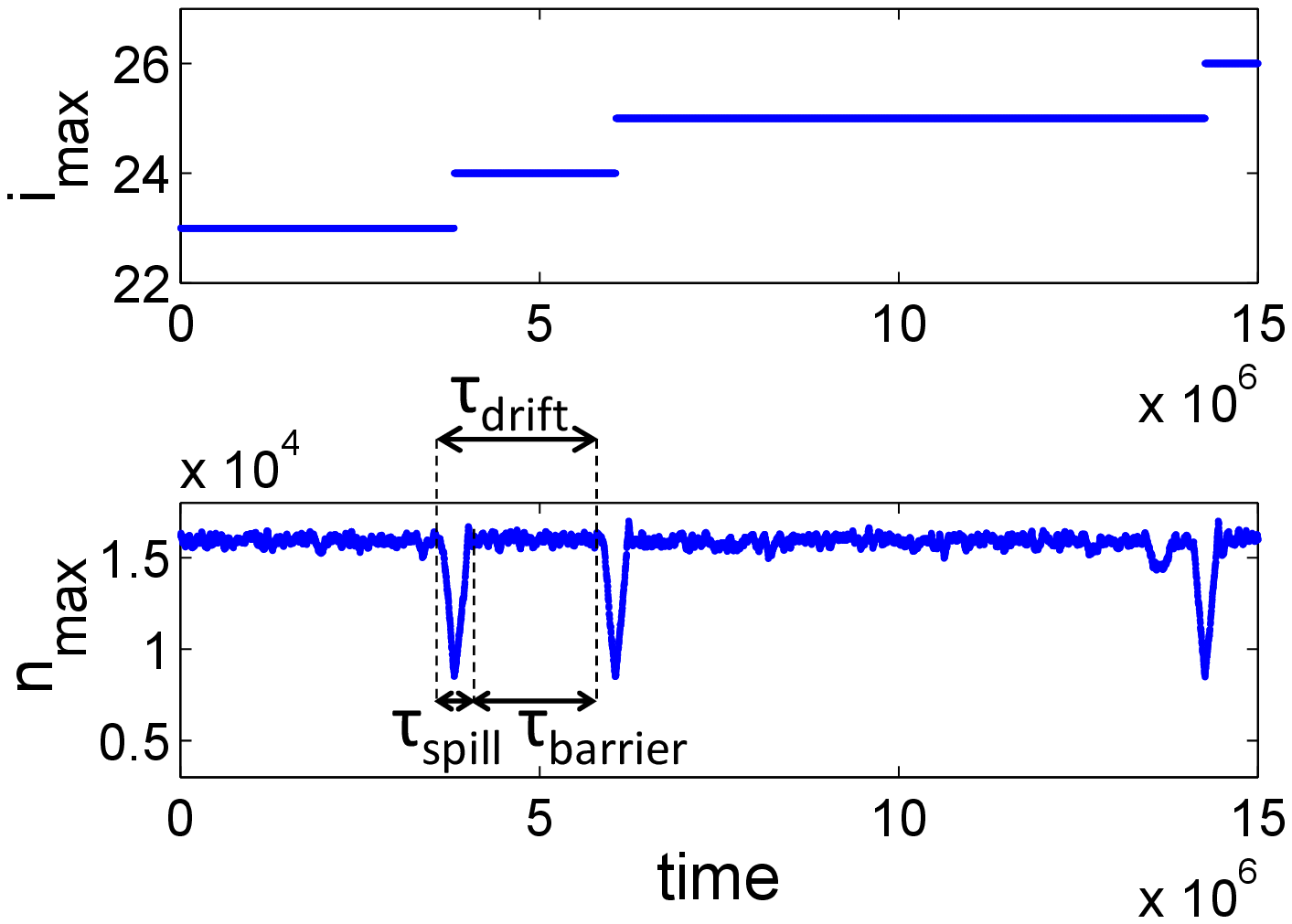}}
  \subfloat[]{\label{fig:weakcondensatepofn}\includegraphics[width=0.49\textwidth]{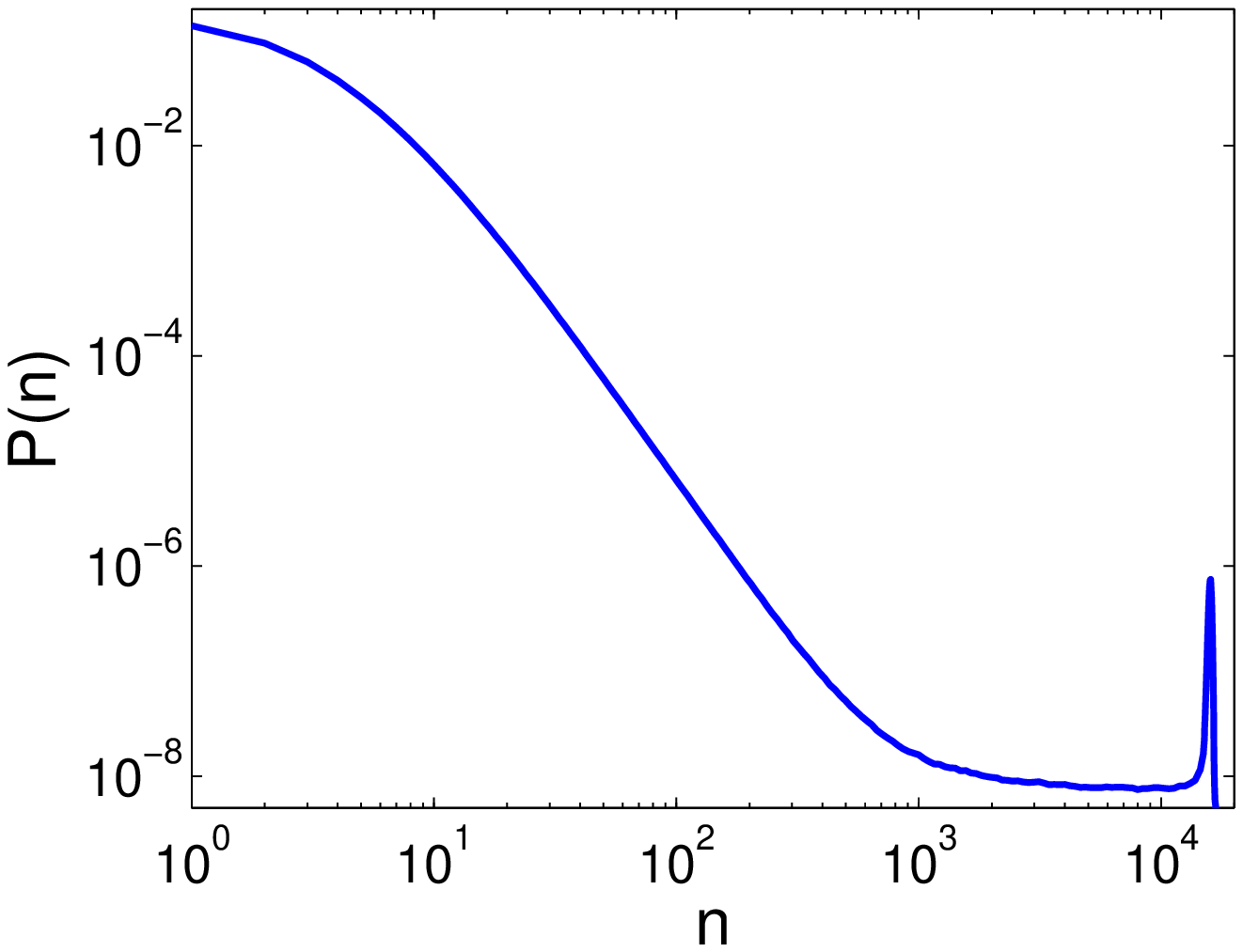}}
  \caption{{(a) The position of the most occupied site
  $i_{\max}$ and its occupation number $n_{\max}$ as a
  function of time, in the weak drift regime, as measured
  on a ring of size $L=2000$ with totally asymmetric hopping,
  $b=5.5$, $\rho = 10$ and $c=2$. The times $\tau_\mysubtext{drift}$,
  $\tau_\mysubtext{spill}$ and $\tau_\mysubtext{barrier}$ are illustrated in
  the bottom panel. (b) The occupation probability $P(n)$
  of a single-site measured in the same system. A plateau is still
  observed, reflecting the slinky motion during transitions for one
  site to the next. However, unlike the strong drift regime, there is
  a peak at high densities reflecting the periods of time when the
  condensate is motionless.
  }}
  \label{fig:weakcondensate}
\end{figure}

All three main features which characterize the strong drift regime
--- a condensate that occupies two sites, its continual drift, and
a plateau in the single site occupation probability --- are modified
in the weak drift regime. In this regime, the condensate occupies a
single site for a long duration of time, and it occupies two sites
only during the (relatively short) time of transition from one site
to the next. This is clearly seen in \fig
\ref{fig:weakcondensatedrfit}, where the occupation and location of
the most occupied site are shown as a function time for a system of
size $L=2000$ with $c=2$ and $b=5.5$ (compare with \fig
\ref{fig:ringcondensatedrfit}, and note the difference in the scale
of the time axes). As a result, a sharp peak is seen in the single
site occupation probability at large values of $n$ (\fig
\ref{fig:weakcondensatepofn}). A plateau is still found at
intermediate values of $n$, but it no longer follows the scaling
relation (\ref{eq:ringpcond}).

\begin{figure}
\center
  \subfloat[]{\label{fig:WeakDriftTaus}\includegraphics[width=0.49\textwidth]{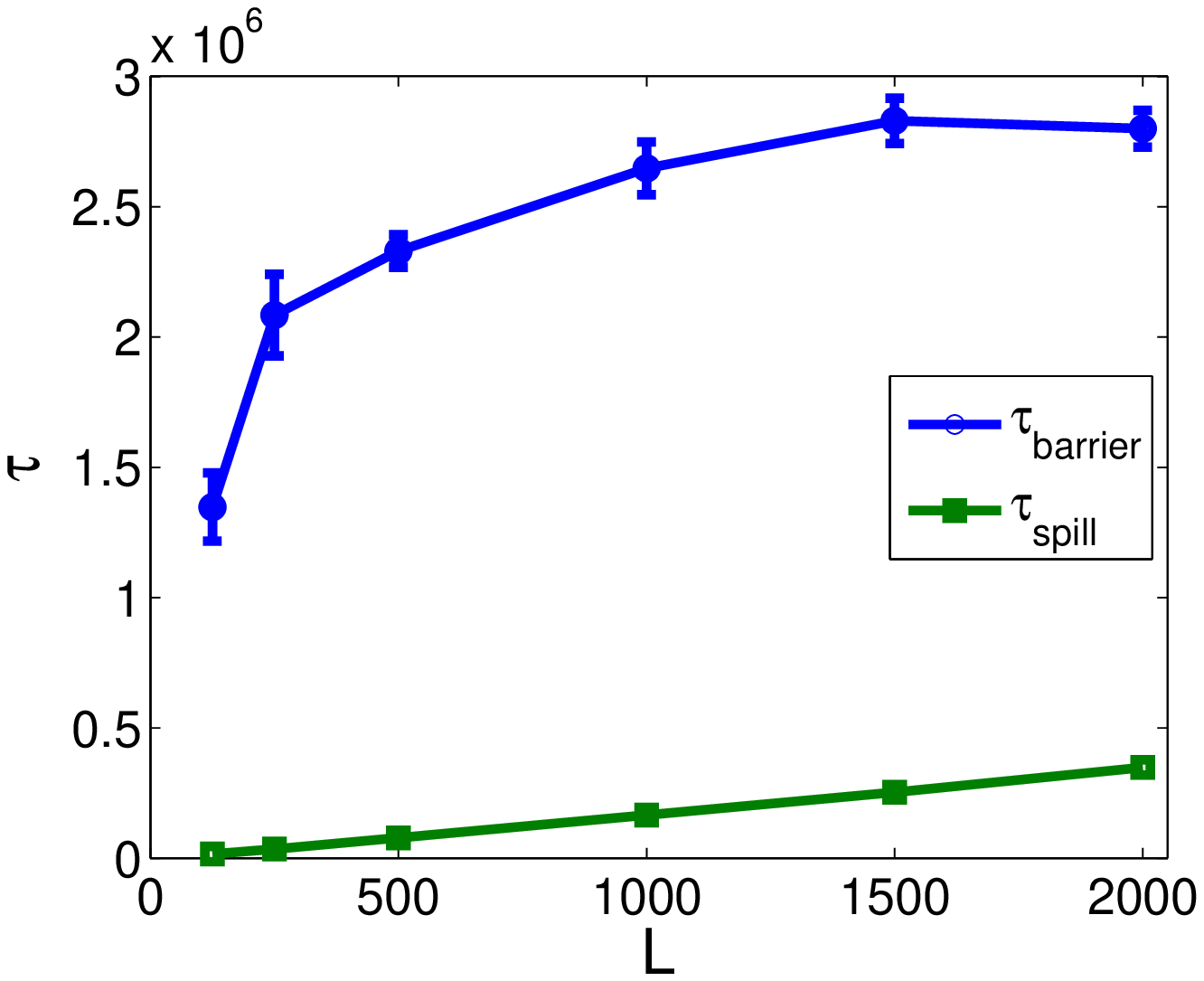}}
  \subfloat[]{\label{fig:WeakDriftTauRatio}\includegraphics[width=0.49\textwidth]{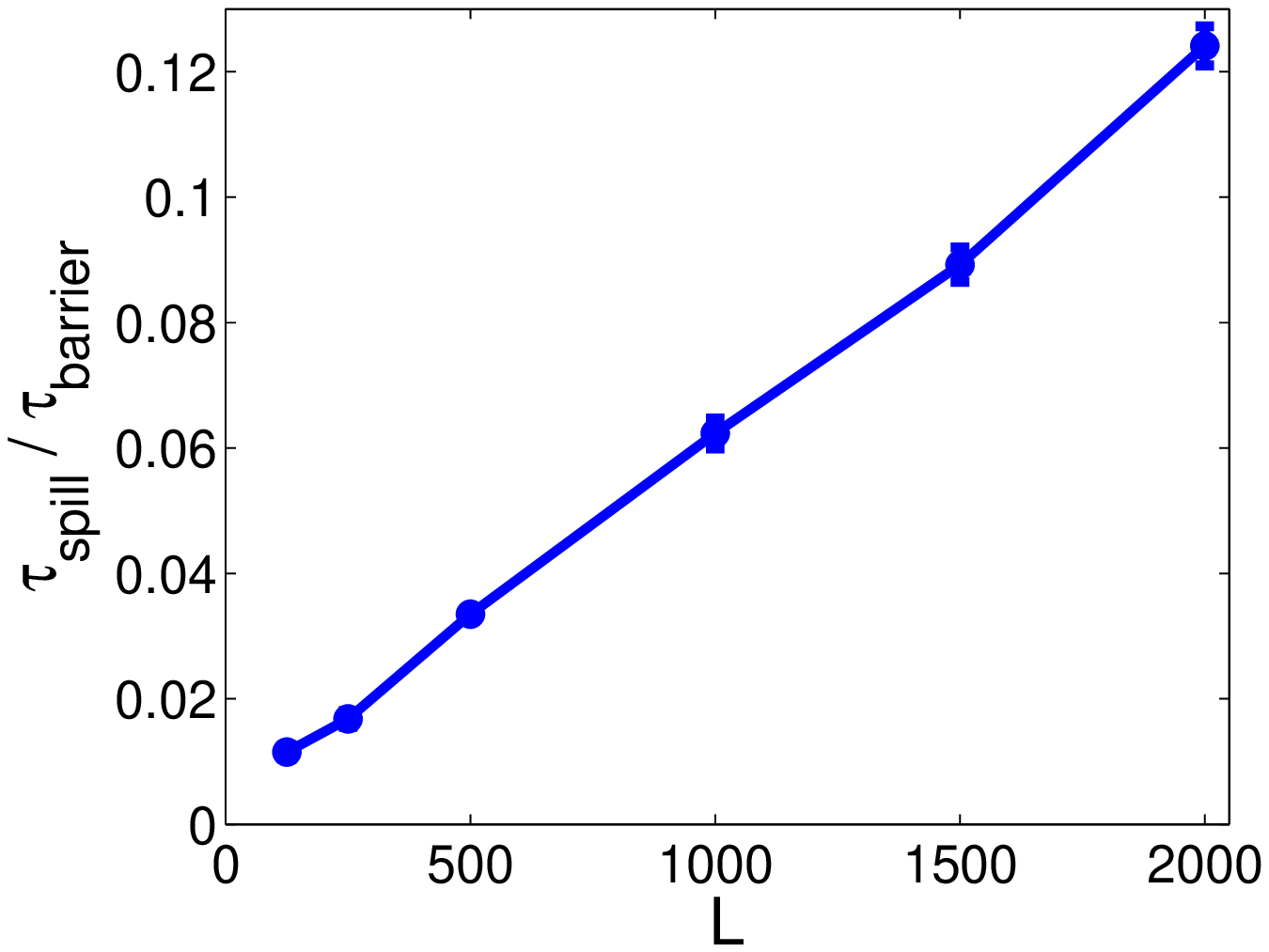}}
  \caption{{(a) $\tau_\mysubtext{barrier}$ and $\tau_\mysubtext{spill}$ (see \fig
  \ref{fig:weakcondensatedrfit}) as a function of the system size.
  (b) The ratio $\tau_\mysubtext{spill} / \tau_\mysubtext{barrier}$ as a function
  of the system size. For the system sizes which we could simulate,
  the condensate motion has not yet converged to its thermodynamic
  limit behavior. If the trend shown here continues at larger $L$,
  eventual condensate motion will be similar to that seen in the strong
  drift regime. The parameters used in the
  simulation are the same as those of \fig \ref{fig:weakcondensate}.
  }}
  \label{fig:weakcondensatesize}
\end{figure}

The characteristics of the weak drift described above suggest that
the condensate is stabilized in one site by a ``barrier''. The
slinky motion is initiated only once fluctuations overcome this
barrier and the number of particles in the condensate decreases
beyond some threshold, or, alternatively, when the number of
particles in the next site increases beyond a threshold.

A possible microscopic mechanism which would give rise to such a
threshold is as follows. Suppose the condensate is located at site
1. As discussed above, the drift motion of the condensate indicates
that the mean current out of site 1 is greater than that leaving
site 2, i.e., $\langle \bar{u}_1(n_1) \rangle > \langle
\bar{u}_2(n_2) \rangle$. However, if $n_2$ is small enough, it might
be that at some moment $\langle \bar{u}_1(n_1) \rangle <
\bar{u}_2(n_2)$. In this case, because $\bar{u}_i(n)$ is a
decreasing function of $n$, particles begin to accumulate in site 2
only after its occupation exceeds a value $n^*$ which is defined by
$\bar{u}_2(n^*) = \langle \bar{u}_1(n_1) \rangle$. Thus, the
condensate begins to spill from site 1 to 2 only after a random
fluctuation brings the occupation of site 2 to $n^*$. If $n^*$ is
large enough, the time until such a fluctuation occurs can be long.
However, this time is expected to remain finite in the thermodynamic
limit $L \to \infty$. If this picture is correct, the erratic motion
of the condensate which characterizes the weak drift regime is
expected to be negligible in the thermodynamic limit, since the time
of the spilling of the condensate scales as the system size $L$. A
more detailed study of such a mechanism for a weak condensate drift
will be presented elsewhere \cite{ToyModel}.

It is not yet known whether this picture provides an accurate
description of the microscopic mechanism which leads to the weak
drift motion. However, numerical evidence indicates that the weak
drift regime may indeed exist only as a finite size effect. To
address this question, we compare the typical time that the
condensate resides on a single site, which we term
$\tau_\mysubtext{barrier}$, with the time it takes the condensate to
``spill'' from one site to the next, which we denote
$\tau_\mysubtext{spill}$. Together, these two time add up to give
the typical time for the drift motion: $\tau_\mysubtext{drift}\equiv
v_\mysubtext{drift}^{-1} = \tau_\mysubtext{spill} +
\tau_\mysubtext{barrier}$, see \fig \ref{fig:weakcondensatedrfit}.
In \fig \ref{fig:WeakDriftTaus}, we present the dependence of
$\tau_\mysubtext{barrier}$ and $\tau_\mysubtext{spill}$ on the
system size. For the system sizes which we were able to study
numerically, $\tau_\mysubtext{spill}$ was seen to grow linearly with
$L$ as expected (it should takes twice as long to move twice as many
particles from one site to the next). However,
$\tau_\mysubtext{barrier}$ is seen to grow slower than linearly.
This trend, which is emphasized when looking at the ratio
$\tau_\mysubtext{spill}/\tau_\mysubtext{barrier}$ (see \fig
\ref{fig:WeakDriftTauRatio}) indicates that although
$\tau_\mysubtext{spill} \ll \tau_\mysubtext{barrier}$ for the system
sizes which were studied, the situation might be reversed at large
enough systems, in which case the motion of the condensate will be
similar to that in the strong drift regime. Whether this trend
continues at larger values of $L$ remains an open question.

\begin{figure}
  \center
  \subfloat[]{\label{fig:weakcondensatemelttimes}\includegraphics[width=0.49\textwidth]{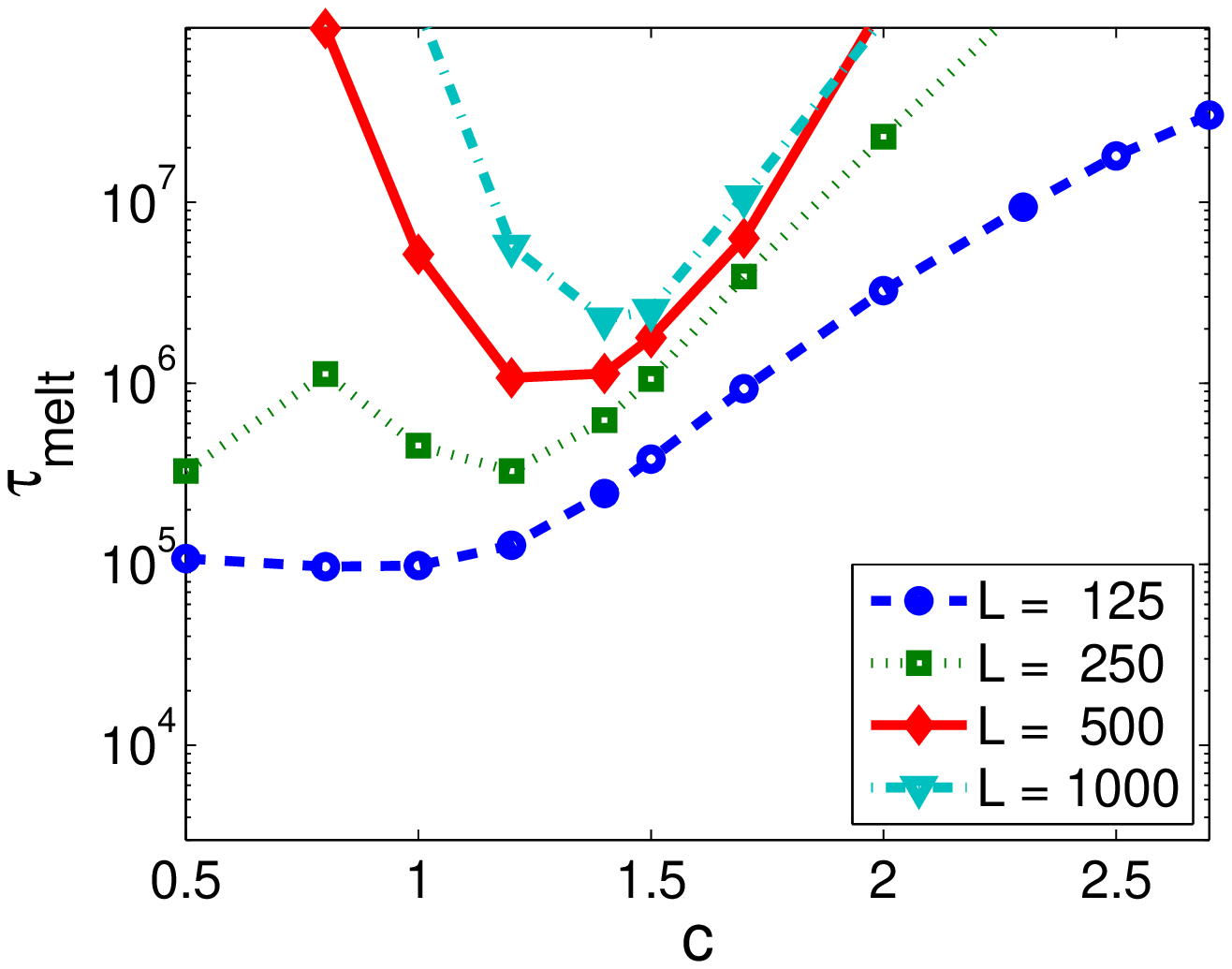}}
  \subfloat[]{\label{fig:weakcondensatedrifttimes}\includegraphics[width=0.49\textwidth]{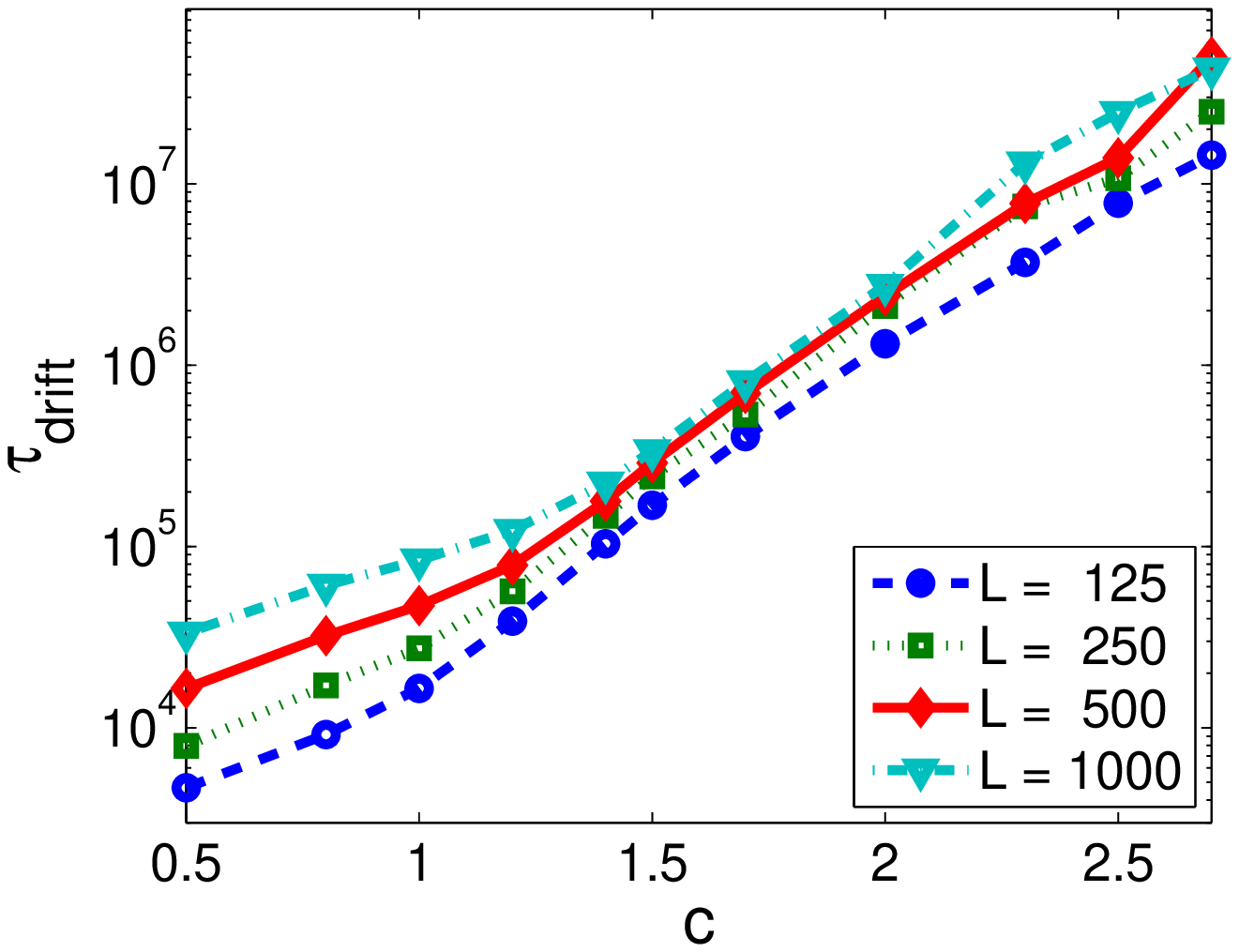}}
  \caption{{The mean time between (a) drift movements of the condensate
  ($\tau_\mysubtext{drift}$) and (b) melting movements
  ($\tau_\mysubtext{melt}$)
  as a function of $c$ for several values of $L$. The entire
  duration of the simulation was $\sim 10^9$ time units, and thus the plots are
  cut off at $\tau \approx 10^8$, beyond which the melting time can
  no longer be estimated reliably with the data available.
  The parameters of the simulation are
  the same as those of \fig \ref{fig:weakcondensate}. Lines are
  guides to the eye.
  }}
\end{figure}

Numerical limitations also hindered the study of the behavior of the
system at the transition between the strong and weak drift regimes,
as well as at higher values of $c$. At values of $1<c<2$, there is a
sharp decrease of the mean time $\tau_\mysubtext{melt}$ between
events at which the condensate melts and reappears in a distant site
(see \fig \ref{fig:weakcondensatemelttimes}). For the values of $L$
which we were able to study these events were still quite frequent,
indicating that the system was still far from thermodynamic
behavior. As $\tau_\mysubtext{drift}$ is seen to grow roughly
exponentially with $c$ (see \fig
\ref{fig:weakcondensatedrifttimes}), when $c$ is larger than about
2, $\tau_\mysubtext{drift}$ becomes comparable with the total length
of the simulation.

\subsubsection{Other types of asymmetric
dynamics}\label{sec:RingHigherD}
The main features of the on-off model, and specifically the drift of
the condensate which was discussed above for the case of
totally-asymmetric hopping, are quite robust to small changes in the
dynamics of the model. We shall now mention a few such modified
models which exhibit a similar behavior in the condensed phase.

We begin with the on-off model with partially asymmetric dynamics,
where each time a particle hops it can jump to the right with
probability $1-p$ or to the left with probability $p$. Totally
asymmetric dynamics corresponds to $p=0$. If an asymmetric system is
in the strong drift regime and $p$ is increased slightly, no
significant changes in its behavior are seen, and in particular it
remains in the strong drift regime. When $p$ is further increased, a
transition to the weak drift regime occurs in the numerical
simulations. This transition is similar to the one discussed above
in the totally-asymmetric case when $c$ is increased beyond 1, and
it too is accompanied by a sharp dip in $\tau_\mysubtext{melt}$. For
a system of size $L=1000$ with $c = 1$ and $b=5.5$ the transition
was found to occur at around $p=0.1$. Beyond this transition, the
drift velocity rapidly decreases as the dynamics approaches the
symmetric dynamics at $p=1/2$ at which point no drift of the
condensate is seen. It should be noted that in the symmetric case,
when the condensate relocates to a different site there seems to be
no preference to its neighboring sites. Rather, the condensate melts
and reappears at a distant site, as in the Markovian case.

A drift of the condensate is also observed when the site is not
turned completely off at $\tau = 0$. Simulations with hopping rates
of the form $u(n,\tau) = u(n)v(\tau)$ with $v(\tau)$ as in \eqn
(\ref{eq:weakstrongrates}), $v_0 = 0.5$ and $b = 5.5$, exhibit
strong drift behavior when $c=0.4$ and weak drift behavior when $c =
1$.

\begin{figure}
\center
  \includegraphics[width=0.6\textwidth]{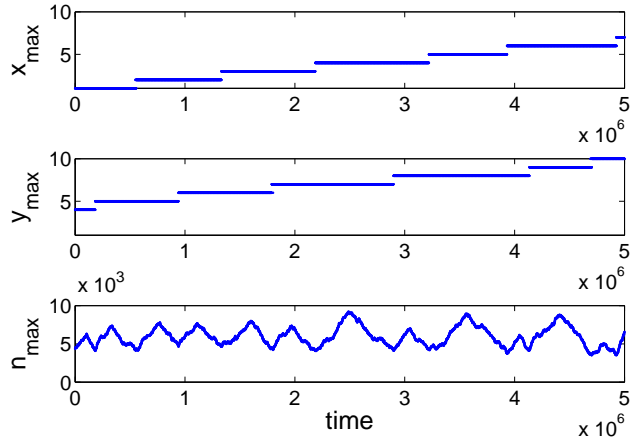}
  \caption{{Drift of the condensate in a 2-dimensional square
  lattice with bias to the up and right directions.
  The top two panels show the $x$- and $y$-components
  $x_\mysubtext{max}$ and $y_\mysubtext{max}$ of the
  location of the most occupied site as a function of time
  and the bottom panel its occupation number $n_\mysubtext{max}$. The lattice
  is of size $20 \times 20$, and the parameters of the
  simulation are $\rho = 30$, $b = 5.5$ and $c = 0.4$. The asymmetry
  of the hopping is as described in the text.
  }}
  \label{fig:2d-drift}
\end{figure}

Condensate drift, both weak and strong, also occurs in 2-dimensional
nearest-neighbor asymmetric on-off models. A particularly
interesting case is when the hopping bias is not parallel to any of
the lattice directions. \fig \ref{fig:2d-drift} displays the motion
of the condensate in a $20 \times 20$ square lattice with periodic
boundary conditions where each time a particle hops it either moves
one site up or one site to the right with equal probabilities. The
figure shows the $x$- and $y$-coordinates of the most occupied site
and its occupation. It is intriguing to notice that the condensate
moves alternatively up and to the right in quite an orderly fashion.
Snapshots of the lattice (not presented here) reveal that the
condensate typically consists of an L-shaped group of three highly
occupied sites. At higher values of $c$ the orderly motion is
destroyed and the condensate drifts in the weak regime.

\section{Exactly solvable non-Markovian ZRP}\label{sec:Exact}

In this section we present a non-Markovian ZRP whose steady-state
probability distribution factorizes into single site terms, similar
to the usual Markovian ZRP. Thus, the steady-state distribution, the
effective hopping rates and $b_\mysubtext{eff}$ can be calculated
exactly. A version of the model with totally asymmetric hopping is
analyzed first in \sect \ref{sec:ExactTA}, and then the model and
results are generalized to symmetric and partially asymmetric
hopping in \sect \ref{sec:ExactSym}.

\subsection{Totally asymmetric dynamics in 1-d}\label{sec:ExactTA}
\subsubsection{Description of the model}
The exactly solvable non-Markovian ZRP is a variant of the on-off
model in which the advance of a clock of a site depends on the clock
states of neighboring sites. Before presenting general results for
partially-asymmetric hopping and for lattices in any dimensions, we
begin for simplicity by considering a one-dimensional lattice with
totally-asymmetric hopping.

The model is similar to the one described above in \sect
\ref{sec:description}: at each site $i$ of a one-dimensional lattice
of $L$ sites there are $n_i$ particles, and a clock variable $\tau_i
=0,1$, signifying ``on'' and ``off''. Note that our notation here
differs from that of \sect \ref{sec:MFonoff}, where $\tau$ was
allowed to take any integer value. This is not a significant
difference, since, as discussed there, identifying all $\tau \geq 1$
clock states with $\tau = 1$ does not affect the dynamics. A
particle can hop from site $i$ to $i+1$ with rates
(\ref{eq:onoffrates}), and once a particle jumps the clock at the
target site is reset to zero. The only difference in the dynamics of
the exactly solvable model is in the way the clocks are updated: the
clock at site $i$ can change from 0 to 1 only if $\tau_{i+1}=1$. The
allowed dynamical moves can be summarized as
\begin{eqnarray}\label{eq:ExactModelDscrp}
\ldots,(n_i,1),(n_{i+1},\tau_{i+1}),\ldots &\myxrightarrow{u(n_i)}&
\ldots, (n_i\!-\!1,1),(n_{i+1}\!+\!1,0), \ldots
\nonumber \\
\ldots, (n_i,0), (n_{i+1},1), \ldots &\myxrightarrow{c}& \ldots,
(n_i,1),(n_{i+1},1), \ldots
\end{eqnarray}
(here $(n_i,\tau_i)$ signifies the occupation and clock state of
site $i$).

\subsubsection{The steady-state distribution}\label{sec:ExactTAsteady}
The goal of this section is to construct the steady state
distribution of this model and show that it has a factorized form.
Before doing so, we note that the factorized form is somewhat
different from that presented in \eqn
(\ref{eq:mfZRPproductmeasure}). The reason for the difference is
that states in which all sites are off cannot be reached by the
dynamics of the model (all other states are possible). This
introduces some correlations between the sites beyond those
generated by the conservation of particles. The product measure
which we discuss below is therefore of the form
\begin{equation}\label{eq:ExactProductMeasure}
\mathcal{P}({\mathbf n},\boldsymbol{\tau}) =
Z^{-1}_{L,N}\prod_{i=1}^L f(n_i,\tau_i)\,\delta\bigg(\!\sum_{i=1}^L
n_i - N\bigg)\bigl[1-\delta(\boldsymbol{\tau})\bigr],
\end{equation}
where $\delta(\boldsymbol{\tau}) = 1$ if $\boldsymbol{\tau} = 0$,
i.e., if all $\tau_i = 0$, and is zero otherwise. Here $f(n,\tau)$
are the single-site weights. The normalization is accordingly given
by
\begin{equation}\label{eq:ExactProductPartition}
Z_{L,N} = \sum_{{\mathbf n},\boldsymbol{\tau}}\prod_{i=1}^L
f(n_i,\tau_i)\,\delta\bigg(\!\sum_{i=1}^L n_i -
N\bigg)\bigl[1-\delta(\boldsymbol{\tau})\bigr].
\end{equation}
In the thermodynamic limit, the weight of configurations with
$\boldsymbol{\tau} = 0$ becomes negligible, and therefore adding the
square-brackets term in (\ref{eq:ExactProductMeasure}) and
(\ref{eq:ExactProductPartition}) does not affect this limit. We
describe some properties of this factorized form in
\ref{sec:AppendixExactFactorized}. Note that the same product form
also describes the finite-size product measure of the mean-field
on-off model which was considered in \sect \ref{sec:MFonoff}.

The dynamics (\ref{eq:ExactModelDscrp}) defines an ergodic process
(on the set of all configurations with a given number of particles
and at least one ``on'' site), and therefore it has a unique steady
state distribution. We now show that this distribution has a
factorized form (\ref{eq:ExactProductMeasure}). This is done by
assuming such a factorized form, and showing that it is indeed the
unique stationary solution of the master equation. We begin by
writing down the master equation. To this end we define a function
$W_{\tau_i,\tau_{i+1}}(n_i,n_{i+1})$ by
\begin{eqnarray}\label{eq:ExactWDef}
\fl \mathcal{P}({\mathbf n},\boldsymbol{\tau})W_{1,1}(n_i,n_{i+1})
&\equiv& \mathcal{P}({\mathbf n},\{\ldots,\tau_{i-1},0,1,\ldots\})c
-
\mathcal{P}({\mathbf n},\boldsymbol{\tau})u(n_i), \nonumber \\
\fl \mathcal{P}({\mathbf n},\boldsymbol{\tau})W_{0,0}(n_i,n_{i+1})
&\equiv& 0, \nonumber \\
\fl \mathcal{P}({\mathbf n},\boldsymbol{\tau})W_{0,1}(n_i,n_{i+1})
&\equiv& -\mathcal{P}({\mathbf n},\boldsymbol{\tau})c,  \\
\fl \mathcal{P}({\mathbf n},\boldsymbol{\tau})W_{1,0}(n_i,n_{i+1})
&\equiv& \nonumber -
\mathcal{P}({\mathbf n},\boldsymbol{\tau})u(n_i) + \sum_{\tau' = 0,1} u(n_i+1) \times \\
\fl & &  \mathcal{P}\Bigl(\{\ldots,n_i+1,n_{i+1}-1,\ldots\},
\{\ldots,\tau_{i-1},1,\tau',\tau_{i+2},\ldots\}\Bigr). \nonumber
\end{eqnarray}
Note that $W_{\tau_i,\tau_{i+1}}(n_i,n_{i+1})$ in fact depends on
the full configuration $({\mathbf n},\boldsymbol{\tau})$. We
suppress this dependence in the notation because if
$\mathcal{P}({\mathbf n},\boldsymbol{\tau})$ has a factorized form,
$W$ indeed depends only on the occupation and clock states of two
adjacent sites (see \eqn (\ref{eq:ExactWDefFactorized}) below).

Using the function $W$, the master equation can be written as
\begin{equation}\label{eq:ExactMaster}
\dot{\mathcal{P}}({\mathbf n},\boldsymbol{\tau}) =
\mathcal{P}({\mathbf n},\boldsymbol{\tau}) \sum_{i=1}^L
W_{\tau_i,\tau_{i+1}}(n_i,n_{i+1}).
\end{equation}
We elucidate \eqns (\ref{eq:ExactWDef}) and (\ref{eq:ExactMaster})
through an example. Consider a configuration of a lattice of 4 sites
with clocks $\boldsymbol{\tau} = \{1,1,0,0\}$ and some occupations
${\mathbf n}$. The different transitions into this configuration and
out of this configuration can be enumerated one bond at a time:
\begin{itemize}
\item Sites 1 and 2: Since both clocks are on, the only possible transition
involving both sites which would lead to this configuration is an
advance of the clock at site 1, which occurs with rate $c$. The only
possible transition out of this configuration which involves the two
sites is a particle hopping from 1 to 2. Therefore, this bond
contributes two terms to the master equation,
\begin{equation}\label{eq:ExactExampleFirst}
\mathcal{P}({\mathbf n},\{0,\tau_2,\ldots\})c - \mathcal{P}({\mathbf
n},\boldsymbol{\tau})u(n_1) = \mathcal{P}({\mathbf
n},\boldsymbol{\tau})W_{1,1}(n_1,n_2).
\end{equation}

\item Sites 2 and 3: The first clock of the two is on while the second is off.
The only possible transition involving these two sites leading to
this configuration is a particle hopping from 2 to 3, and this is
also the only possible transition out of this configuration.
Therefore, this bond contributes to the master equation
\begin{eqnarray}
\fl \sum_{\tau_3 = 0,1}
\mathcal{P}\Bigl(\{\ldots,n_2+1,n_3-1,\ldots\},\{\ldots,1,\tau_3,\ldots\}\Bigr)
u(n_2+1) - \mathcal{P}({\mathbf n},\boldsymbol{\tau})u(n_2) = \nonumber \\
= \mathcal{P}({\mathbf n},\boldsymbol{\tau})W_{1,0}(n_2,n_3).
\end{eqnarray}

\item Sites 3 and 4: Both clocks are off, and therefore no
transitions which involve only this bond are possible. One can
define the contribution to the master equation as
\begin{equation}
0 = \mathcal{P}({\mathbf n},\boldsymbol{\tau})W_{0,0}(n_3,n_4).
\end{equation}

\item Sites 4 and 1: The first clock is off and the second is on.
There are no transitions involving only this bond which can lead to
this configuration. However, there is a possible transition out of
this configuration, by an advance of the clock of site 4. The
contribution from this bond is therefore
\begin{equation}\label{eq:ExactExampleLast}
-\mathcal{P}({\mathbf n},\boldsymbol{\tau})c = \mathcal{P}({\mathbf
n},\boldsymbol{\tau})W_{0,1}(n_4,n_1).
\end{equation}
\end{itemize}
Summing up \eqns
(\ref{eq:ExactExampleFirst})--(\ref{eq:ExactExampleLast}) leads to
the master equation (\ref{eq:ExactMaster}). A similar analysis shows
that the master equation has exactly the same form for any
configuration and for any lattice size.

Now assume that $\mathcal{P}({\mathbf n},\boldsymbol{\tau})$ has the
factorized form (\ref{eq:ExactProductMeasure}). In this case, the
definition (\ref{eq:ExactWDef}) has the simpler form
\begin{eqnarray}\label{eq:ExactWDefFactorized}
W_{1,1}(n_i,n_{i+1}) &=&
\frac{f_\mysubtext{off}(n_i)}{f_\mysubtext{on}(n_{i})} \, c -
u(n_i),
\nonumber \\
W_{0,0}(n_i,n_{i+1}) &=& 0, \nonumber \\
W_{0,1}(n_i,n_{i+1}) &=& -c, \nonumber \\
W_{1,0}(n_i,n_{i+1}) &=& \frac{f_\mysubtext{on}(n_i+1)f(n_{i+1}-1)}
{f_\mysubtext{on}(n_i)f_\mysubtext{off}(n_{i+1})} \, u(n_i+1) -
u(n_i),
\end{eqnarray}
where $f_\mysubtext{off}(n) \equiv f(n,0)$,  $f_\mysubtext{on}(n)
\equiv f(n,1)$, and $f(n) \equiv \sum_{\tau'}f(n,\tau')$. In the
steady state, the left-hand side of \eqn (\ref{eq:ExactMaster})
vanishes and the equation becomes $\sum_{i=1}^L
W_{\tau_i,\tau_{i+1}}(n_i,n_{i+1}) = 0$. This equation is solved by
explicitly constructing its unique solution. This is done in two
steps. First, we show that if one finds $f(n,\tau)$ which satisfies
\begin{eqnarray}
0 &=& W_{1,1}(n_i,n_{i+1}) \label{eq:ExactPairwiseBalance1} \\
0 &=& W_{1,0}(n_i,n_{i+1}) + W_{0,1}(n_j,n_{j+1})
\label{eq:ExactPairwiseBalance2}
\end{eqnarray}
for any $n_i, n_{i+1},n_j,n_{j+1}$, this $f$ is a solution to the
equation. Then, we construct such an $f$.

The first step is achieved by noting that the number of
$W_{1,0}(n_i,n_{i+1})$ terms in the sum (\ref{eq:ExactMaster})
exactly equals the number of $W_{0,1}(n_j,n_{j+1})$ terms in the
sum, since any configuration of $\tau_i$'s must have the same number
of $01$ and $10$ nearest-neighbor pairs. Therefore, all terms in the
sum (\ref{eq:ExactMaster}) vanish either individually or in pairs,
and the sum equals zero. We now construct a solution $f$ which
satisfies
(\ref{eq:ExactPairwiseBalance1})--(\ref{eq:ExactPairwiseBalance2}).
Condition (\ref{eq:ExactPairwiseBalance1}) is equivalent to
\begin{equation}\label{eq:ExactStMaster1}
c f_\mysubtext{off}(n) = u(n) f_\mysubtext{on}(n).
\end{equation}
For condition (\ref{eq:ExactPairwiseBalance2}), note that
$W_{0,1}(n_j,n_{j+1})$ is in fact independent of $n_j,n_{j+1}$ (see
\eqn (\ref{eq:ExactWDefFactorized})). Therefore, this condition
together with (\ref{eq:ExactStMaster1}) yield
\begin{equation}
\frac{f(n_i + 1)}{f(n_i)}\,\frac{u(n_i +1)} {c + u(n_i+1)} =
\frac{f(n_{i+1})}{f(n_{i+1}-1)}\, \frac{u(n_{i+1})}{c +u(n_{i+1})},
\end{equation}
As the occupations $n_i$ and $n_{i+1}$ may vary independently, this
equation holds only if both sided are equal to a constant, which
might be set to $1/c$ without loss of generality (as it only affects
the normalization constant $Z_{L,N}$). We therefore find that
\begin{equation}\label{eq:ExactStMaster2}
f(n)\bar{u}(n) = f(n-1)
\end{equation}
where
\begin{equation}\label{eq:ExactUbar}
\frac{1}{\bar{u}(n)} = \frac{1}{c} + \frac{1}{u(n)},
\end{equation}
compare with equations (\ref{eq:simplerec}) and
(\ref{eq:onoffubar}). Choosing the constant to be $1/c$ guarantees
that, as we show below, $\bar{u}(n)$ as defined in \eqn
(\ref{eq:ExactUbar}) are the effective hopping rates.

The conclusion from \eqns
(\ref{eq:ExactStMaster1})--(\ref{eq:ExactUbar}) is that the
factorized probability distribution of the form
(\ref{eq:ExactProductMeasure}) with
\begin{eqnarray}
f(n,1) &\equiv& f_\mysubtext{on}(n) = \frac{c}{c+u(n)}\,f(n), \label{eq:ExactPon}\\
f(n,0) &\equiv& f_\mysubtext{off}(n) = \frac{u(n)}{c+u(n)}\,f(n), \label{eq:ExactPoff}\\
f(n) &=& \prod_{k=1}^n \frac{1}{\bar{u}(k)} = \prod_{k=1}^n \biggl[
\frac{1}{c} + \frac{1}{u(n)} \biggr]. \label{eq:ExactPtotal}
\end{eqnarray}
is the stationary solution of model. It is easy to verify using
(\ref{eq:ExactUbar}) and (\ref{eq:ExactPon}) that $\bar{u}(n) =
\frac{f_\mysubtext{on}(n)u(n)}{f(n)}$, and therefore $\bar{u}(n)$
are the effective hopping rates as defined in \eqn (\ref{eq:ubar}).

Using the results (\ref{eq:ExactPon})--(\ref{eq:ExactPtotal}), one
can calculate numerically the stationary probability for any
configuration in a finite system of size $L$ with $N$ particles
(recursion relations that facilitate this calculation are presented
in \ref{sec:AppendixExactFactorized}). Condensation in the model is
determined by the probability measure in the thermodynamic limit.
The factorized product measure (\ref{eq:ExactProductMeasure}) and
\eqn (\ref{eq:ExactStMaster2}) imply that this model has the same
thermodynamic behavior as a Markovian ZRP with effective hopping
rates (\ref{eq:ExactUbar}) (see \ref{sec:AppendixExactFactorized}).
One can thus study condensation in the model using known properties
of the ZRP, as was done in \sect \ref{sec:MeanField}. In particular,
for rates of the form $u(n) = 1 + b/n + O(n^{-2})$, one finds
\begin{equation}
\bar{u}(n) = \frac{c}{1+c}\,\frac{1+\frac{b}{n} +
O(n^{-2})}{1+\frac{b}{(1+c)n} + O(n^{-2})} = J_c
\Bigl[1+\frac{b_\mysubtext{eff}}{n}+O(n^{-2})\Bigr]
\end{equation}
with
\begin{equation}\label{eq:ExactJcAndBeff}
J_c = \frac{c}{1+c}<1 \quad \mytext{and} \quad b_\mysubtext{eff} =
J_c b
 < b.
\end{equation}
Here, $J_c$ is the current at the critical density, and
$b_\mysubtext{eff}$ is the parameter controlling condensation. In
other words, condensation may occur only when $b_\mysubtext{eff} >
2$, or $b>2/J_c$ (compare with the MF values in \eqns
(\ref{eq:onoffubarfinal})--(\ref{eq:mfbeff1}), and note also that at
criticality, the probability to find a site in the off state is
$P_\mysubtext{off} = J_c/c$, rather than (\ref{eq:poff}) of the MF
model).

\subsection{Partially asymmetric dynamics and higher-dimensional
lattices}\label{sec:ExactSym}
\subsubsection{Description of the model}
The exactly solvable model described above can still be fully
analyzed when the dynamics is generalized to partially asymmetric or
symmetric dynamics and to certain higher dimensional lattices.
Moreover, the stationary distribution turns out the be independent
on the asymmetry or the dimension. We described the generalized
dynamics and its solution in this section.

First, consider dynamics on a 1-d lattice that allows for partially
asymmetric hopping. This is implemented as discussed above in \eqn
(\ref{eq:ZRPringdynamics}): when a particle jumps from an ``on''
site $i$ (an event which occurs with a rate $u(n)$), it randomly
chooses its target site: with probability $1-p$ it moves to site
$i+1$ and otherwise (i.e. with probability $p$) it moves to $i-1$.
Here $0\leq p\leq1$ is the asymmetry parameter: $p=1/2$ corresponds
to symmetric dynamics, while $p=0$ corresponds to a totally
asymmetric bias to the right.

For the stationary distribution to factorize, one must also modify
the update rule for the clock variable, in the following manner. At
each ``off'' site, an attempt to update the clock is made with rate
$c$. Once an attempt is made at, say, site $i$, a neighboring site
is chosen at random with the \emph{same} asymmetry parameter $p$:
site $i+1$ is chosen with probability $1-p$ and site $i-1$ with
probability $p$. Finally, if the chosen site is in an ``on'' state,
the clock of site $i$ is turned on. The generalized dynamics can be
summarized as
\begin{eqnarray}\label{eq:ExactAsymModelDscrp}
\ldots,(n_i,1),(n_{i+1},\tau_{i+1}),\ldots
&\myxrightarrow{(1-p)u(n_i)}& \ldots,
(n_i\!-\!1,1),(n_{i+1}\!+\!1,0), \ldots
\nonumber \\
\ldots,(n_{i-1},\tau_{i-1}),(n_i,1),\ldots
&\myxrightarrow[\phantom{(1-p)u(n_i)}]{p u(n_i)}& \ldots,
(n_{i-1}\!+\!1,0),(n_i\!-\!1,1), \ldots \nonumber \\
\ldots, (n_i,0), (n_{i+1},1), \ldots
&\myxrightarrow[\phantom{(1-p)u(n_i)}]{(1-p)c}& \ldots,
(n_i,1),(n_{i+1},1), \ldots \nonumber \\
\ldots, (n_{i-1},1), (n_i,0), \ldots
&\myxrightarrow[\phantom{(1-p)u(n_i)}]{p c}& \ldots, (n_{i-1},1),
(n_i,1), \ldots,
\end{eqnarray}
where the first line describes a particle jump from $i$ to the
right, the second describes a jump to the left, and the third and
fourth lines describe the two update processes of the clock at site
$i$.

In a similar fashion, the model can be generalized to symmetric or
biased dynamics on higher dimensional lattices. Here we consider for
concreteness cubic lattices in $d$-dimensions, although the argument
which we present below for the factorization of the stationary
distribution is valid for other lattices, e.g. a triangular lattice
in $2d$.\footnote{Note, however, that the argument does not hold for
\emph{all} higher dimensional lattices. For example, the argument
fails for a $2d$ honeycomb lattice.} As in the partially asymmetric
case, a particle leaves any site $i$, if it is on, with rate $u(n)$.
It then selects its target from among the $2d$ nearest neighbors of
$i$ according to an asymmetry probability vector $p_a$, where
$a=1,\ldots,2d$ denotes the direction (for example, in two
dimensions $a=1,2,3,4$ could correspond to north, east, south and
west) and $\sum_{a=1}^{2d} p_a = 1$. A choice of $p_a = 1/2d$ for
all $a$ corresponds to symmetric dynamics, and any other choice
would result in biased hopping. The clock update rule in the
$d$-dimensional case is similarly generalized: if site $i$ is off,
an attempt to update its clock is made with rate $c$. At each
attempt, the neighbor of $i$ in the direction $a$ is chosen with
probability $p_a$, and if the chosen neighbor is on the clock of $i$
is updated.

Since 1-dimensional partially asymmetric hopping is a particular
case of $d$-dimensional dynamics, both cases are treated below
together.

\subsubsection{The steady-state distribution}
The factorization in the generalized case is demonstrated as done
above, by explicitly constructing the stationary measure. To this
end it is once again assumed that the stationary measure has the
factorized form
(\ref{eq:ExactProductMeasure})--(\ref{eq:ExactProductPartition}).
The master equation for this factorized distribution reads, at the
steady state,
\begin{equation}\label{eq:ExactAsymMaster}
0 = \dot{\mathcal{P}}({\mathbf n},\boldsymbol{\tau}) =
\mathcal{P}({\mathbf n},\boldsymbol{\tau}) \sum_{a=1}^{2d} p_a
\biggl[\sum_{i=1}^L W_{\tau_i,\tau_{i+a}}(n_i,n_{i+a})\biggr],
\end{equation}
where site $i+a$ denotes the neighbor of site $i$ in the direction
$a$, and $W$ is defined in (\ref{eq:ExactWDefFactorized}).

The key observation which facilitates finding a solution to this
master equation is that for each $i$ and $a$ such that $\tau_i = 1$
and $\tau_{i+a} = 0$, there exists exactly one site $j$ whose clock
is $\tau_j = 0$ while $\tau_{j+a} = 1$. This can be seen for example
by examining the clocks of all sites on the ray which starts at site
$i$ and is in direction $a$ (i.e., by examining sites $i+2a, i+3a,
\ldots$), which leads to a situation similar to the one-dimensional
case. Therefore, a solution to \eqn (\ref{eq:ExactAsymMaster}) can
be found if $W_{1,1}(n,n') = 0$ and $W_{1,0}(n,n') + W_{0,1} = 0$
for all $n$ and $n'$ (note again that $W_{0,1} \equiv W_{0,1}(n,n')$
is independent of $n,n'$). These are precisely the conditions which
appeared in the totally asymmetric case, and therefore they are
fulfilled by the same solution --- \eqns
(\ref{eq:ExactPon})--(\ref{eq:ExactPtotal}).

We have thus shown that the stationary distribution of the
generalized model factorizes, and moreover it is independent of
asymmetry and lattice dimension. In particular, condensation is
independent of the asymmetry parameter, and the results of \sect
\ref{sec:ExactTAsteady} apply.

\section{Conclusions}\label{sec:conclusion}

The analysis presented above reveals that non-Markovian dynamics may
have two major effects on the condensation transition of the ZRP.
First, the parameter $b$ which controls condensation is
``renormalized'' by the existence of memory in the dynamics, and
thus a memory may suppress or induce condensation. For models with
mean-field dynamics and for an exactly solvable variant of the
model, the effective rates could be computed exactly, and thus the
modified criterion for condensation was found. Numerically, the
condensation in models with nearest-neighbor hopping were also found
to be controlled by an effective $b$, although one which differs
from the mean-field value. Calculating the effective hopping rates
in nearest-neighbor models remains an open problem which may be of
practical importance when one wishes to use a non-Markovian ZRP to
study condensation in other systems.

A second effect of the memory is perhaps more dramatic: the
condensate is found to move from one site to the next when the
dynamics is of asymmetric nearest-neighbor hopping. Numerical
studies of finite systems identify two modes of condensate drift: a
strong-drift regime with continuous ``slinky'' motion and a
weak-drift regime in which the motion is more erratic. Both modes of
motion are rather robust to changes in the dynamics. The behavior of
the model in the thermodynamic limit is not yet known, and it would
be interesting to ascertain whether there is a sharp transition
between them, or, if such a transition does not exist, to understand
the crossover from one regime to the other.

The mechanism which leads to the condensate drift is understood on a
heuristic level and is expected to be a generic feature of many
systems which undergo a condensation transition and which are
asymmetric and have some spatial correlations \cite{ToyModel}.
However, a more quantitative understanding of this drift, for
example the calculation of the drift velocity, remains an important
open problem. It is also interesting to explore similar effects in
other mass-transport systems, such as driven diffusive systems and
shaken granular gases. In this respect, it should be noted that a
mass-transport model with a moving condensate was recently
identified in \cite{BartekEvans2012Explosive}. There, a variant of
the ZRP is studied which has a factorized steady-state and in which
unbound hopping rates lead to a condensate which reaches an infinite
velocity. A product measure steady state with a moving condensate is
not possible in systems with finite hopping rates like ours.

We have also studied an exactly soluble variant of the non-Markovian
model with nearest-neighbor hopping whose steady state factorizes.
In this variant, as in the mean-field model, condensation is
controlled by an effective $b$ and no condensate motion appears. It
should be notes that although the model has a product measure,
particle currents are temporally correlated.

\ack

The support of the Israel Science Foundation (ISF) is gratefully
acknowledged.


\appendix

\section{Regular and irregular clocks}\label{appendix:RegClocks}
As stressed above, the internal clock variables $\tau_i$ do not
measure an exact time, but rather proceed in an irregular stochastic
fashion. In this Appendix, it is shown that regular clock, that
proceed in a deterministic continuous fashion, may be obtained from
the dynamical rules (\ref{eq:modeldscrp}) by taking an appropriate
limit.

We denote the clock variables in this Appendix as $m_i$ instead of
$\tau_i$, to emphasized that they may attain only integer values. In
order to obtain regular clocks, define new clock variables $\tau_i =
m_i d\tau$, where $d\tau$ is an infinitesimal time unit which will
eventually be taken to zero. The new clock variables are no longer
integer: they can attain any value $\tau = 0, d\tau,
2d\tau,3d\tau,\ldots$, and in the limit of infinitesimal $d\tau$
they become continuous variables. In addition, the rate with which
$m_i$ advances to $m_{i}+1$ is taken as $c=1/d\tau$. Finally, the
hopping rates out of each site $i$ are taken to depend on $\tau_i$
rather than $m_i$, and thus they can be written as $u(n_i,\tau_i)$.
The limit of regular clocks is then obtained by taking the limit
$d\tau \to 0$, while keeping $\tau_i$ fixed.

For example, consider an on-off model with regular clocks, whose
hopping rates are $u(n,\tau) = u(n)\Theta(\tau/\tau_0-1)$, where
$\tau_0$ is a constant and $\Theta(x)$ is the Heaviside theta
function. Such a model may be achieved by considering
irregular-clock models (\ref{eq:modeldscrp}) with rates $u(n,m) =
u(n)\Theta(m\, d\tau/\tau_0-1)$ and $c=1/d\tau$, and taking the
limit $d\tau \to 0$ while keeping the constant $\tau_0$ fixed. In
this regular-clock on-off model, whenever a particle hops into a
site this site is turned off for a duration of exactly $\tau_0$ time
units. The solution of such a model with mean-field dynamics may be
found from an analysis similar to that presented in \sect
\ref{sec:MeanField} \cite{HirschbergThesis}. Similarly, more general
regular-clock models with rates $u(n,\tau) = u(n)v(\tau)$ can be
obtained by taking the limit of irregular-clock models with rates
$u(n,m) = u(n)v( m\, d\tau/\tau_0)$ (the function $v$ remains
unchanged when taking the limit).


\section{Properties of factorized distributions of the
form (\ref{eq:ExactProductMeasure})}\label{sec:AppendixExactFactorized}

In this Appendix we present some of the properties of the stationary
distribution of the exactly solvable on-off model, which has the
factorized form (\ref{eq:ExactProductMeasure}) with partition
function (\ref{eq:ExactProductPartition}). The goal of this Appendix
is to present recursion relations which allow the calculation of
this product measure for any finite system size, and to demonstrate
that such product measures lead to the same thermodynamic behavior
as (\ref{eq:mfZRPproductmeasure}) and (\ref{eq:mfZRPpartition}).

We begin by defining two auxiliary partition sums,
\begin{eqnarray}
\bar{Z}_{L,N} &=& \sum_{{\mathbf n},\boldsymbol{\tau}}\prod_{i=1}^L
f(n_i,\tau_i)\,\delta\bigg(\!\sum_{i=1}^L n_i - N\bigg), \\
Z_{L,N}^\mysubtext{off} &=& \sum_{{\mathbf n}}\prod_{i=1}^L
f_0(n_i)\,\delta\bigg(\!\sum_{i=1}^L n_i - N\bigg),
\end{eqnarray}
and two auxiliary distributions,
\begin{eqnarray}
\bar{\mathcal{P}}({\mathbf n},\boldsymbol{\tau}) &=&
\frac{1}{\bar{Z}_{L,N}} \prod_{i=1}^L
f(n_i,\tau_i)\,\delta\bigg(\!\sum_{i=1}^L n_i - N\bigg), \\
\mathcal{P}^\mysubtext{off}({\mathbf n}) &=& \frac{1}
{Z_{L,N}^\mysubtext{off}} \prod_{i=1}^L
f_0(n_i)\,\delta\bigg(\!\sum_{i=1}^L n_i - N\bigg).
\label{eq:RecursionPOff}
\end{eqnarray}
Here and in the rest of this section we denote $f_0(n) \equiv
f(n,0)$ and $f_1(n) \equiv f(n,1)$. This is done to avoid confusion
with the superscript ``off'', which will be used below to denote
quantities calculated using the distribution
(\ref{eq:RecursionPOff}). As before, we denote $f(n) \equiv f_0(n) +
f_1(n)$, and we adopt the convention of \sect \ref{sec:Exact}
whereby the clocks may have only two values, $\tau = 0,1$.

Using these notations and the definition
(\ref{eq:ExactProductPartition}) one immediately finds that
\begin{equation}\label{eq:RecursionZRelation}
Z_{L,N} = \bar{Z}_{L,N} - Z_{L,N}^\mysubtext{off}.
\end{equation}
We first analyze the auxiliary distributions before treating the
original problem. By summing over the occupations and clock states
of all sites but one, the probability to find a single site in any
given state is found to be
\begin{eqnarray}\label{eq:RecursionAuxPofn}
\fl \bar{P}(n,\tau) &\equiv& \sum_{{ n_2,\ldots,n_L}\atop{
\tau_2,\ldots,\tau_L}}
\bar{\mathcal{P}}(n,n_2,\ldots,n_L;\tau,\tau_2,\ldots,\tau_L)
\delta\bigg(\!\sum_{i=2}^L n_i - (N-n)\bigg) = \nonumber \\
\fl &=& f(n,\tau)\frac{\bar{Z}_{L-1,N-n}}{\bar{Z}_{L,N}}, \\
\fl P^\mysubtext{off}(n) &\equiv& \sum_{n_2,\ldots,n_L}
\mathcal{P}^\mysubtext{off}(n,n_2,\ldots,n_L)
\delta\bigg(\!\sum_{i=2}^L n_i - (N-n)\bigg) =
f_0(n)\frac{Z^\mysubtext{off}_{L-1,N-n}}{Z^\mysubtext{off}_{L,N}}.
\end{eqnarray}
Summing both equations over $n$ leads to the recursion relations
\begin{eqnarray}
\bar{Z}_{L,N} &=& \sum_{n=0}^N  \bar{Z}_{L-1,N-n} f(n), \\
Z^\mysubtext{off}_{L,N} &=& \sum_{n=0}^N
Z^\mysubtext{off}_{L-1,N-n} f_0(n).
\end{eqnarray}
When $f(n,\tau)$ are known, these recursion formulas can be used for
a numerical calculation of the auxiliary partition sums, and thus,
using (\ref{eq:RecursionZRelation}) also of $Z_{L,N}$.

Knowing the partition function $Z_{L,N}$, other quantities of
interest can be computed. For example, repeating the calculation of
(\ref{eq:RecursionAuxPofn}) for the original product measure yields
\begin{equation}\label{eq:RecursionPofn}
P(n,\tau) = f(n,\tau) \frac{Z_{L-1,N-n} + \delta_{\tau,1}
Z^\mysubtext{off}_{L-1,N-n}}{Z_{L,N}},
\end{equation}
where $\delta_{\tau,1}$ is the Kronecker delta. The current can be
found in a similar fashion by calculating
\begin{eqnarray}
\fl \langle u(n) \rangle  \equiv \sum_{n=1}^N P(n,1)u(n) =
\sum_{n=1}^N f(n,1)u(n) \frac{\bar{Z}_{L-1,N-n}}{Z_{L,N}} =
\sum_{n=0}^{N-1} f(n)
\frac{\bar{Z}_{L-1,N-1-n}}{Z_{L,N}} \nonumber \\
= \frac{\bar{Z}_{L,N-1}}{Z_{L,N}},
\end{eqnarray}
where (\ref{eq:ExactPon})--(\ref{eq:ExactPtotal}) were used to
deduce that $f(n,1)u(n) = f(n-1)$, and we have used
(\ref{eq:RecursionZRelation}) and (\ref{eq:RecursionPofn}).

In the thermodynamic limit, the partition function can be analyzed
by transforming to the grand-canonical ensemble. Mathematically this
is done by introducing the grand-canonical partition function which
is the generating function
\begin{equation}\label{eq:RecursionGCZdef}
\mathcal{Z}_L(z) \equiv \sum_{N=0}^\infty z^N Z_{L,N}.
\end{equation}
Using the definition (\ref{eq:ExactProductPartition}), one can split
the sum into two contributions, $\mathcal{Z}_L(z) =
\bar{\mathcal{Z}_L} - \mathcal{Z}^\mysubtext{off}_L$, where
\begin{equation}
\bar{\mathcal{Z}}_L \equiv \sum_{N=0}^\infty z^N \bar{Z}_{L,N} =
\bar{F}(z)^L, \qquad \mathcal{Z}^\mysubtext{off}_L \equiv
\sum_{N=0}^\infty z^N Z^\mysubtext{off}_{L,N} =
F^\mysubtext{off}(z)^L,
\end{equation}
and
\begin{equation}\label{eq:RecursionFBar}
\bar{F}(z) \equiv \sum_{n=0}^\infty f(n) z^n, \qquad
F^\mysubtext{off}(z) \equiv \sum_{n=0}^\infty f_0(n) z^n.
\end{equation}
Using (\ref{eq:ExactPon})--(\ref{eq:ExactPtotal}), one has
$F^\mysubtext{off}(z) = z\bar{F}(z)/c$, from which the
grand-canonical partition function is found to be
\begin{equation}\label{eq:RecursionGCZ}
\mathcal{Z}_L(z) = \bar{\mathcal{Z}}_L(z)
\biggl[1-\Bigl(\frac{z}{c}\Bigr)^L\biggr].
\end{equation}
If the radius of convergence of the sum (\ref{eq:RecursionGCZdef}),
or equivalently of (\ref{eq:RecursionFBar}), is smaller than $c$,
then the correction due to the weak correlation between clocks is
exponentially small when $L$ is large. For hopping rates of the form
$u(n) \simeq 1 + b/n$, this radius of convergence is $z_c = c/(1+c)
< c$ (see \eqn (\ref{eq:ExactPtotal})), and therefore, in this case
$(z/c)^L$ is indeed negligible. Note that $z_c$ is the current $J_c$
of the canonical system at the condensation transition, \eqn
(\ref{eq:ExactJcAndBeff}).

The relation between the fugacity $z$ and the canonical density
$\rho$ is given by
\begin{equation}\label{eq:RecursionGCdensity}
\rho = \frac{z}{L}\,\frac{\partial \log \mathcal{Z}(z)}{\partial z}
= z\frac{\bar{F}'(z)}{\bar{F}(z)} + \frac{1}{1-(\frac{z}{c})^{-L}},
\end{equation}
which is an implicit equation for $z(\rho)$. This is the same
expression as that of a Markovian ZRP with rates $\bar{u}(n)$, up to
a correction which is exponentially small in $L$.



\bibliographystyle{unsrt}
\bibliography{longpaperbib}

\end{document}